\documentclass[preprint,showpacs,preprintnumbers,amsmath,amssymb,superscriptaddress, nofootinbib]{revtex4}
\usepackage{graphicx,color}
\usepackage{amsmath,amssymb}
\usepackage{url}
\usepackage{epstopdf}

\newcommand{\be}{\begin{equation}}
\newcommand{\ee}{\end{equation}}
\newcommand{\bea}{\begin{eqnarray}}
\newcommand{\eea}{\end{eqnarray}}

\newcommand{\crn}{\nonumber \\}

\newcommand{\bc}{\begin{center}}
	\newcommand{\ec}{\end{center}}

\newcommand{\De}{\Delta}

\newcommand {\ba}{\begin{array}}
	\newcommand {\ea}{\end{array}}
\newcommand{\ben}{\begin{enumerate}}
	\newcommand{\een}{\end{enumerate}}
	
\usepackage{epsfig,graphicx}
\usepackage{bm}
\usepackage{dcolumn}
\begin{document}
	\title{General one-loop formulas for decay $h\rightarrow Z\gamma$}
	\author{L.~T.~ Hue} \email{lthue@.iop.vast.ac.vn}
	\affiliation{Institute of Physics, Vietnam Academy of Science and Technology, 10 Dao Tan, Ba Dinh, Hanoi, Vietnam}
	\affiliation{Bogoliubov Laboratory for Theoretical Physics,
		Joint Institute for Nuclear Researches, Dubna, Russia}
	\author{A.~B.~Arbuzov} \email{arbuzov@theor.jinr.ru}
	\affiliation{Bogoliubov Laboratory for Theoretical Physics,
		Joint Institute for Nuclear Researches, Dubna, Russia}
\affiliation{Department of Higher Mathematics, Dubna State University, Dubna, Russia}
	\author{T.~T.~Hong} \email{tthong@agu.edu.vn}
	\affiliation{Department of Physics, An Giang University, Ung Van Khiem Street, Long Xuyen, An Giang, Vietnam}
	\affiliation{Department of Physics, Hanoi Pedagogical University 2, Phuc Yen, Vinh Phuc, Vietnam}
	\author{T.~Phong~Nguyen} \email{thanhphong@ctu.edu.vn}
	\affiliation{Department of Physics, Cantho University, 3/2 Street, Ninh Kieu, Cantho, Vietnam}
	\author{D.~T.~Si} \email{dangtrungsi@cantho.edu.vn}
	\affiliation{Department of Education and Training of Can Tho City, No 39, 3/2 Street, Can Tho, Vietnam}
	\author{H.~N.~Long} \email{hoangngoclong@tdtu.edu.vn}
	\affiliation{Theoretical Particle Physics and Cosmology Research Group, Advanced Institute of Materials Science, Ton Duc Thang University, Ho Chi Minh City, Vietnam}
	\affiliation{Faculty of Applied Sciences,
		Ton Duc Thang University, Ho Chi Minh City, Vietnam}
	\begin{abstract}
	Radiative corrections to the decay $h\rightarrow Z\gamma$ are evaluated in the one-loop approximation. The unitary gauge is used. The analytic  result is expressed in terms of the Passarino-Veltman functions. The calculations are applicable for the Standard Model as well for a wide class of its gauge extensions. In particular, the decay width of a charged Higgs boson $H^{\pm}\rightarrow W^{\pm}\gamma$ can be derived. The consistence of our formulas and several specific earlier results is shown.
	\end{abstract}
	\pacs{
		{14.80.Bn,
			12.60.Cn,
			12.15.Ji,
			12.15.Lk
		}
	}
	\maketitle
	\allowdisplaybreaks
\section{Introduction}

After the discovery of the Higgs boson particle at LHC in 2012~\cite{Htn,Htn1},
many improved measurements confirmed the consistence of its quantum numbers and couplings
with the Standard Model (SM) predictions, including the loop-induced coupling
$h\gamma\gamma$~\cite{Khachatryan:2016vau,HiggsExconstrain1}.  Meanwhile, another
loop-induced coupling $hZ\gamma$ related to the decay $h\rightarrow Z\gamma$ has not
been measured yet even so that the predicted decay rate is of the same order
as the one of $h\rightarrow \gamma\gamma$ in the SM case~\cite{hzgapredict}.
The partial decay width $h\rightarrow Z\gamma$ was calculated within the SM framework
and its supersymmetric extension~\cite{hzga1,hzga2,Bardin:1991dp,hzgamssm,HHunter,Cao:2013ur,Belanger:2014roa,Hammad:2015eca}. From the experimental side, this decay channel is now been searched at the LHC by both
CMS and ATLAS collaborations~\cite{hzgaex1,hzgaex2, hzgaex3}. Many discussions concerning
studies of this channel are going also in planned experimental projects as at the LHC
as well as at future $e^+e^-$ and even 100~TeV proton-proton
colliders~\cite{hzgaexim1, hzgaexim2}. While the effective coupling  $h\gamma\gamma$ is now very
strictly constrained experimentally, the coupling $hZ\gamma$ might be still significantly
different from the SM prediction in  certain SM extensions  because of the
 $Z$ boson couplings with new particles. Studies the decay of the SM-like Higgs boson
$h\rightarrow Z\gamma$ affected by the presence of new fermions and charged scalars
were performed in several models beyond the SM (BSM) having the same SM gauge
group~\cite{hzgamssm, HzgaTHD, GMmodel, hdcay1,Dev:2013ff}.

At the one loop level, the amplitude of the decay $h\rightarrow Z\gamma$ contains also
contributions from new gauge boson loops of the BSM models constructed from larger
electroweak gauge groups such as the left-right (LR), 3-3-1, and 3-4-1 models~\cite{lr1, lr2, lr3, lr4, lr5, lr6, g331, g331a, g331b, g331c, g331d, g331e, g331f,g331g, g341,g341a}.
Calculating these contributions is rather difficult in the usual 't~Hooft-Feynman gauge,
because of the appearance of many unphysical states, namely Goldstone bosons and ghosts
which always exist along with the gauge bosons. They create a very large number of Feynman
diagrams.  In addition, their couplings are indeed model dependent, therefore it is hard
to construct general formulas determining vector loop contributions using
the t'~Hooft-Feynman gauge. This problem has been mentioned recently~\cite{GMmodel}
in a discussion of the Georgi-Machacek model, where only new Higgs multiplets are added
to the SM. The reason is that the new Higgs bosons will change the couplings of unphysical
states with the gauge bosons $Z$ and $W^\pm$. In the left-right models predicting new gauge bosons that contribute to the amplitude of the decay $h\rightarrow Z\gamma$, previous calculations in this gauge  were also model dependent \cite{hzgalr1,hzgalr2}.  An approach introduced recently in Ref. \cite{Goodsell:2017pdq} for calculating the  decay  $h\rightarrow Z\gamma$,  with the help of numerical computation packages,  may be more convenient.

The technical difficulties caused by unphysical states will vanish if calculations have
been done in the unitary gauge. There the number of Feynman diagrams as well as the
number of necessary couplings become minimum, namely only those which contain physical
states are needed. Then the Lorentz structures of these couplings are well defined,
and hence the general analytic formulas of one-loop contributions from gauge boson loops
can be constructed. But in the unitary gauge we face complicated forms of the gauge boson
propagators, which generate many dangerous divergent terms. Fortunately, many of them are
excluded by the condition of on-shell photon in the decay $h\rightarrow Z\gamma$.
The remaining ones will vanish systematically when loop integrals are written in terms
of the Passarino-Veltman (PV) functions \cite{PVfunc}. This situation will be demonstrated
in this work explicitly.
Moreover, the choice of the unitary gauge allows us to derive general analytic formulas
for one-loop contributions involving various gauge bosons to the amplitude of the decay
$h\rightarrow Z\gamma$. The formulas will be given in terms of standard PV functions
defined by ref.~\cite{PVDenner} and in the LoopTools library~\cite{LoopTools}.
The analytic forms of these PV functions are also presented so that our results can
be compared with the earlier results   calculated independently  in specific cases.
In addition, the analytic formulas can be implemented into numerical stand-alone packages
without dependence on the LoopTools. Our results can be translated into the general analytic
form used to calculate the amplitudes of the charged Higgs decay
$H^{\pm}\rightarrow W^{\pm}\gamma$ which is also an interesting channel predicted
in many BSM models. Our results can be easily compared also with those given recently
in~\cite{GMmodel}, which were calculated in the 't~Hooft-Feynman gauge. Moreover, our
results can be cross-checked with another one-loop formula expressing new gauge boson
contributions in the gauge-Higgs unification (GHU) model~\cite{hzgaGHU}.

The decay $H\rightarrow Z\gamma$ of the  new heavy neutral Higgs boson $H$ in the SM supersymmetric model was also mentioned in \cite{Belanger:2014roa}. The signal strength of this decay was shown to be very sensitive with the parameters  of the model, hence it may give interesting information on the parameters once it is detected. Many other BSM also contain heavy neutral Higgs bosons $H$, and  the one loop amplitudes of their  decays $H\rightarrow Z\gamma$ may include many significant contributions that do not appear in the case of the SM-like Higgs boson. Some of the complicated contributions are usually ignored by qualitative estimations. The analytic formulas introduced in this work are enough to determine more quantitatively these approximations.

Apart from the above BSM with non-Abelian gauge group extensions, there are BSM with additional Abelian gauge groups \cite{Holdom:1985ag,Babu:1996vt}. These models predict new kinetic mixing parameters between Abelian gauge bosons, which appear in the couplings of the neutral physical gauge bosons including the SM-like one, for example see \cite{Cassel:2009pu}.  Our calculation in the unitary gauge are also applicable with  only condition that  couplings of physical states are determined.

Our paper is organized as follows. Section~\ref{Feyrule} will give the general
notations and Feynman rules necessary for calculation of the width of the decay
$h\rightarrow Z\gamma$ in the unitary gauge. In section~\ref{analytic} we present
important steps of the derivation of the analytic formula for the total contribution
of gauge boson loops. We also introduce all other one-loop contributions from
possible new scalars and fermions appearing in BSM models. In section~\ref{comparision},
the comparison between our results with previous ones will be discussed, including the
case of charged Higgs decays. We will emphasize the contributions from gauge boson loops
both in decays of neutral CP-even and charged Higgs bosons. Our result will be applied to discuss on two particular models in  section~\ref{application}. In Conclusions we will highlight
important points obtained in this work. In the first  Appendix, we review notations of the PV functions given by LoopTools and their analytic forms used in other
popular numerical packages. Two other Appendices contain detailed calculations of the one-loop fermion contributions to the amplitude $h\to Z\gamma$ and the relevant couplings in the LR models discussed in our work.

\section{\label{Feyrule} Feynman diagrams and rules}
\begin{figure}[t]
	\centering
	\includegraphics[width=15cm]{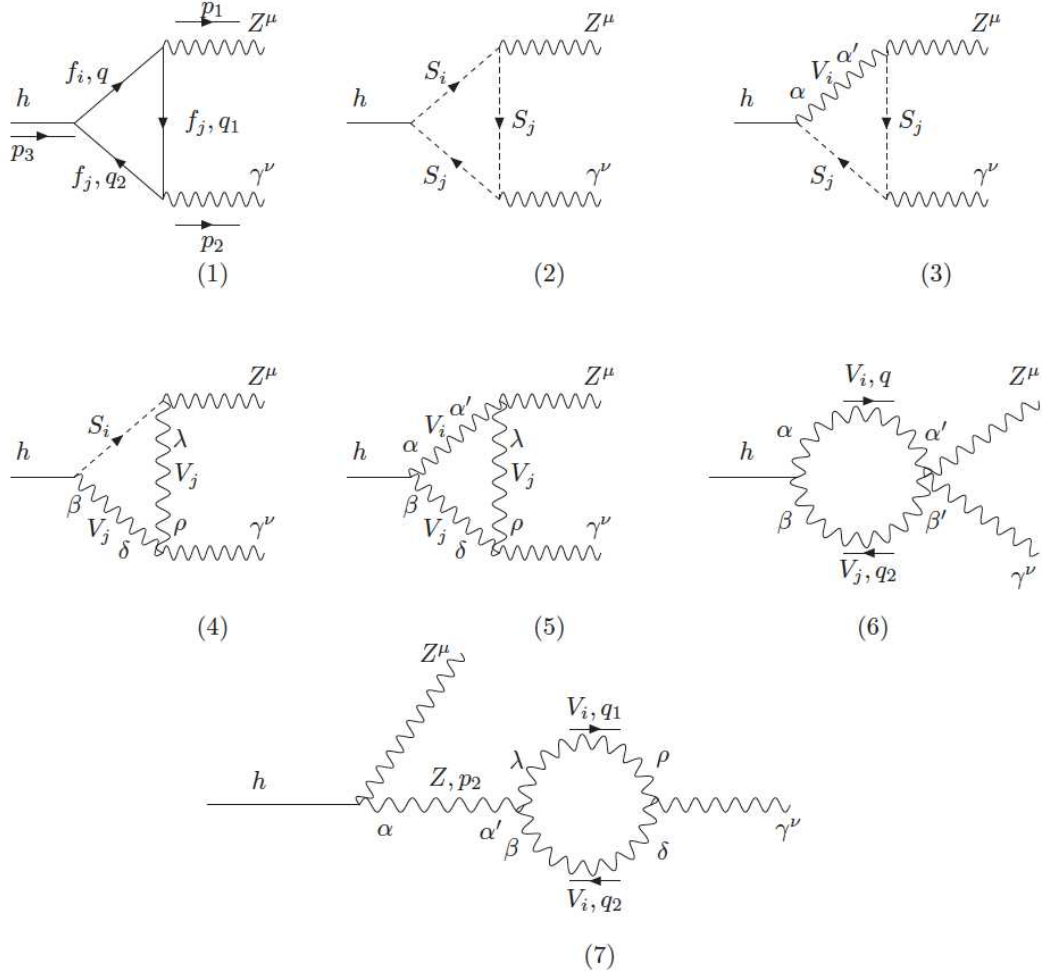}\\
	\caption{One-loop diagrams contributing to the decay  $h\rightarrow Z\gamma$, where $f_{i,j}$, $S_{i,j}$  and $V_{i,j}$ are fermions, Higgs, and gauge bosons, respectively.}\label{loopHd}
\end{figure}

The amplitude of the decay $h\rightarrow Z\gamma$ is generally defined as
\begin{align}
\mathcal{M}(h\rightarrow Z\gamma)&\equiv \mathcal{M}\left(Z_{\mu}(p_1),\gamma_{\nu}(p_2),h(p_3)\right) \varepsilon^{\mu*}_1(p_1) \varepsilon^{\nu*}_2 (p_2)\crn
&\equiv \mathcal{M}_{\mu\nu}\varepsilon^{\mu*}_1 \varepsilon^{\nu*}_2,
\label{GenAm}
\end{align}
where $\varepsilon^{\mu}_1$ and $\varepsilon^{\nu}_2$ are the polarization vectors of
the $Z$ boson and the photon $\gamma$, respectively. The external momenta $p_{1}$, $p_{2}$,
and $p_{3}$ satisfy the condition $p_3=p_1+p_2$ with the directions denoted in
figure~\ref{loopHd} where one-loop Feynman diagrams contributing to the decay are presented.
Only diagrams which are relevant in the unitary gauge are mentioned. The on-shell conditions
are $p_1^2=m_Z^2$, $p_2^2=0$, and $p_3^2=m_h^2$.

The decay amplitude is generally written in the following form~\cite{hzgamssm}:
\begin{align}
\mathcal{M}_{\mu\nu}\equiv  F_{00}\, g_{\mu\nu}+ \sum_{i,j=1}^2F_{ij} p_{i\mu}p_{j\nu}+ F_{5}\times i\epsilon_{\mu\nu\alpha\beta} p_{1}^{\alpha}p_{2}^{\beta}, \label{mmunu}
\end{align}
where $\epsilon_{\mu\nu\alpha\beta}$ is the totally antisymmetric tensor
with $\epsilon_{0123}=-1$ and $\epsilon^{0123}=+1$~\cite{peskin}.

The equality $\varepsilon^{\nu*}_2 p_{2\nu}=0$ for the external photon implies that
$F_{12,22}$ do not contribute to the total amplitude (\ref{GenAm}). In addition,
the $\mathcal{M}_{\mu\nu}$ in eq.~(\ref{mmunu}) satisfies the Ward identity,
$p_{2}^{\nu}\mathcal{M}_{\mu\nu}=0$, resulting in $F_{11}=0$ and~\cite{hzgamssm}
 \be
 F_{00}=- (p_1.p_2) F_{21}=\frac{(m_Z^2-m_h^2)}{2}F_{21}.  \label{Wardconsequence}
 \ee
 Hence the amplitude (\ref{GenAm}) can be calculated through the form~(\ref{mmunu})
via the following relations
 \begin{align}
 \mathcal{M}(h\rightarrow Z\gamma)&=  \mathcal{M}_{\mu\nu}\varepsilon^{\mu*}_1 \varepsilon^{\nu*}_2,\crn
 \mathcal{M}_{\mu\nu}&= F_{21}\left[-(p_2.p_1) g_{\mu\nu} +p_{2\mu}p_{1\nu}\right]+ F_{5}\times i\epsilon_{\mu\nu\alpha\beta} p_{1}^{\alpha}p_{2}^{\beta}.\label{amp1}
 \end{align}
 The partial decay width then can be presented in the form~\cite{HHunter,GMmodel}
 \be
 \Gamma(h\rightarrow  Z\gamma)=\frac{m_h^3}{32\pi}
 \times \left(1-\frac{m_Z^2}{m_h^2}\right)^3\left(|F_{21}|^2 +|F_5|^2\right).
 \label{GaHZga1}
 \ee
The above formula shows us that we need to find only two scalar coefficients $F_{21}$
and $F_5$ in eq.~(\ref{amp1}). Because $F_5$ arises from only chiral fermion loops,
it is enough to pay attention to terms proportional to $F_{21}p_{2\mu}p_{1\nu}$ for gauge
boson loops. Therefore calculations will be simplified, especially in the unitary gauge.
Combining with notations of the PV functions~\cite{PVfunc}, we will determine explicitly
which terms give contributions to $F_{21}p_{2\mu}p_{1\nu}$, and hence exclude step by step
irrelevant terms throughout our calculations.

\begin{figure}[t]
	\label{counterVer}
	\includegraphics[width=12cm]{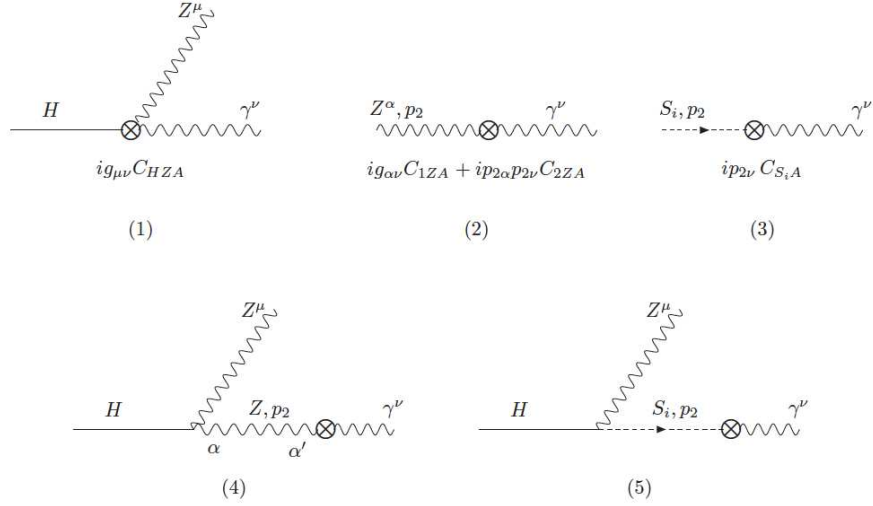}\\
	\caption{Counterterm vertices and related one-loop diagrams contributing to the one-loop amplitude of the decay  $h\rightarrow Z\gamma$.}
\end{figure}

Calculation of the factor $F_{21}$ is very interesting because it does not receive
contributions from diagrams which contain counterterm vertices. The Lorentz structures
of the counterterm vertices are shown in figure~\ref{counterVer}.
The first line represents three additional counterterm vertices. The second line
shows two more diagrams. The total amplitude is the sum of three diagrams 1, 4, and 5
in figure~\ref{counterVer} and all diagrams shown in figure~\ref{loopHd}.
We can see in figure~\ref{counterVer} that, the first diagram contributes only to $F_{00}$.
In the unitary gauge, the propagator of a gauge boson is
\be
\De^{\mu\nu}(k^2,m^2) =\frac{-i}{k^2-m^2}\left(g^{\mu \nu}-\frac{k^{\mu}k^{\nu}}{m^2} \right).
\label{prounitary}
\ee
The Lorentz structures of the two remaining counterterms are
\begin{align*}
i\mathcal{M}^{\mathrm{CT}}_{(4)\mu\nu}&\sim g_{\mu\alpha}\times \left( g^{\alpha\alpha'}- \frac{p_{2}^{\alpha} p_{2}^{\alpha'}}{m_Z^2}\right) \times \left( g_{\alpha'\nu} C_{1ZA} +p_{2\alpha'}p_{2\nu}C_{2ZA} \right)\crn
&= g_{\mu\nu} C_{1ZA} + p_{2\mu}p_{2\nu} \left( C_{2ZA}- \frac{C_{1ZA}}{m_Z^2}\right),\crn
i\mathcal{M}^{\mathrm{CT}}_{(5)\mu\nu}&\sim (p_3 +p_2)_{\mu} \times \left(p_{2\nu}C_{S_iA}\right)= (p_1 +2 p_{2})_{\mu} p_{2\nu} C_{S_iA},
\end{align*}
which contribute only to $F_{00}$, $F_{12}$, and $F_{22}$. The result for the Lorentz
structures is unchanged if the virtual gauge boson $Z$ in diagram~4 is replaced with
the new ones in a gauge extended  versions of the SM. As the result, $F_{21}$ is not
affected by counterterms, therefore we do not need to include them in our calculation.
In addition, $F_{21}$ is finite without including the related counterterm diagrams.
A similar situation in two Higgs doublet models was discussed in~\cite{HzgaTHD}.
Examples for Lorentz structures of the counterterms were given also,
e.g., in refs.~\cite{PVDenner,grace}.

\begin{table}[t]
\begin{tabular}{|c|c|}
\hline
Vertex& Coupling \\
\hline
$h\overline{f_i}f_j$&  $-i\left(Y_{hf_{ij}L}P_L + Y_{hf_{ij}R}P_R\right)$\\
\hline
$hS^{Q}_iS^{-Q}_j$, $hS^{-Q}_iS^{Q}_j$& $-i\lambda_{hS_{ij}}$, $-i\lambda^*_{hS_{ij}}$\\
\hline
$h(p_0) S^{-Q}_i(p_-)V^{Q\mu }_j $, $h(p_0)S^{Q}_i(p_+)V^{-Q\mu}_j$ &  $ig_{hS_iV_j} (p_0-p_{-})_{\mu}$,  $-ig^*_{hS_iV_j} (p_0-p_{+})_{\mu}$  \\
\hline
$h V^{-Q\mu }_iV^{Q\nu}_j$, $h Z^{\mu }Z^{\nu}$ &  $ig_{hV_{ij}} g_{\mu\nu}$, $ig_{hZZ} g_{\mu\nu}$  \\
\hline
$A^{\mu}\overline{f_i}f_i$, $A^{\mu}S^{Q}_iS^{-Q}_i$& $ie\,Q\gamma_{\mu}$, $ie\,Q(p_{+}-p_{-})_{\mu}$\\
\hline
$A^{\mu}(p_0) V^{ Q\nu}_i(p_+)V^{-Q\lambda}_i(p_-)$&  $-ie Q\Gamma_{\mu\nu\lambda}(p_{0}, p_+,p_-)$  \\
\hline
$Z^{\mu}\overline{f_i}f_j$
& $i\left(g_{Zf_{ij}L}\gamma_{\mu}P_L+ g_{Zf_{ij}R}\gamma_{\mu} P_R\right)$\\
\hline
$Z^\mu S^{Q}_i(p_+)S^{-Q}_j(p_-)$& $ig_{ZS_{ij}}(p_{+}-p_{-})_{\mu}$\\
\hline
$ Z^{\mu }V^{Q\nu }_iS^{-Q}_j$, $Z^{\mu}V^{-Q\nu}_iS^{Q}_j $&  $ig_{ZV_iS_j}\, g_{\mu\nu}$, $ig^*_{ZV_iS_j}\, g_{\mu\nu}$  \\
\hline
$Z^{\mu}(p_0) V^{Q\nu}_i(p_+)V^{-Q\lambda }_j(p_-)$&  $-ig_{Z V_{ij}}\Gamma_{\mu\nu\lambda}(p_{0}, p_+,p_-)$  \\
\hline
$Z^{\mu} A^{\nu } V^{Q\alpha}_iV^{-Q\beta}_j$&  $-ie\,Q\,g_{ZV_{ij}}\left(2 g_{\mu\nu}g_{\alpha\beta} - g_{\mu\alpha}g_{\nu\beta}- g_{\mu\beta}g_{\nu\alpha}\right)$  \\
\hline
\end{tabular}
\caption{Couplings involving the decay of CP even neutral Higgs  $h\rightarrow Z\gamma$, in the unitary gauge. A new notation is  $\Gamma_{\mu\nu\lambda}(p_{0}, p_+,p_-)\equiv (p_0-p_+)_{\lambda} g_{\mu\nu} +(p_+-p_-)_{\mu} g_{\nu\lambda} +(p_--p_0)_{\nu} g_{\lambda\mu}$,  where all momenta are incoming, and  $p_{0,\pm}$  are respective momenta of $h$ and charged gauge and Higgs bosons with electric charges $\pm Q$, denoted as  $V_{i,j}^{\pm Q}$ and $S_{i,j}^{\pm Q}$, respectively. The general case of four-gauge-boson coupling is $(2,-1,-1)\rightarrow (a_1,a_2,a_3)$ and $g_{Z\gamma V_{ij}}\neq e\,Q\,g_{ZV_{ij}}$.}\label{tVcoupling}
\end{table}

The Feynman rules used in our calculations are listed in table~\ref{tVcoupling}.
We found them to appear commonly in many gauge extensions of the SM, for example
in the models constructed from the following electroweak gauge symmetries:
$SU(2)_1\times SU(2)_2\times U(1)_Y$, $SU(2)_L\times SU(2)_R\times U(1)_Y$, and
$SU(3)_L\times U(1)_X$~\cite{gaugExtent221, Roitgrund:2014zka, gaugExtent221b, gaugExtent331,gaugExtent331a,C0f}, where an important relation $g_{Z\gamma V_{ij}}=e\,Q\,g_{ZV_{ij}}$ is valid.
It results in that many complicated terms containing dangerous divergences in two
contributions from diagrams 5 and 6 in figure~\ref{loopHd} cancel each other out.

Following LoopTools~\cite{LoopTools}, figure~\ref{loopHd} defines three internal momenta
$q,q_1,q_2$ as follows
\begin{eqnarray}
\label{02}
q_1=q+k_1=q-p_1, \quad
q_2=q+k_2=q-(p_1+p_2),\quad p_1=-k_1, \quad
p_2=k_1-k_2.
\end{eqnarray}
 Our formulas will be written in terms of common well-defined PV functions.
Moreover, we can compare our results with previous works, as well as we can
perform numerical estimates with the help of the LoopTools library.
Definitions and notations for the PV functions are shown in Appendix~\ref{Looptoolnote}.

As the result, we only need to calculate the coefficient $F_{21}$. In the next section,
we will present important steps of how to get contributions from pure gauge boson loops
to $F_{21}$.

\section{\label{analytic} Analytic formulas}

\subsection{Total contribution from diagrams with pure gauge boson mediations}
Here we will consider calculation of the contribution from pure gauge boson loops to
the decay amplitude of $h\rightarrow Z\gamma$. All of them were performed using
the FORM language~\cite{form1,form2}. Other contributions from diagrams which contain
only one or two internal gauge boson lines are computed more easily.

The contribution from diagram~5 from figure~\ref{loopHd} reads
\begin{align}
\label{M5Vij}
i \mathcal{M}_{(5)\mu \nu} &= 2 \times \int \frac{d^dq}{(2 \pi)^d} (ig_{hV_{ij}}\,g_{\alpha \beta}) \frac{-i}{D_0} \left( g^{\alpha \alpha'} - \frac{q^\alpha q^{\alpha'}}{m_1^2} \right) \times \left[-ig_{ZV_{ij}}\Gamma_{\mu\alpha'\lambda}(-p_1,q,-q_1)\right] \crn
&\times \frac{-i}{D_1} \left( g^{\lambda \rho} - \frac{q_1^\lambda q_1^{\rho}}{m_{2}^2} \right)\times \left[-ie\,Q\, \Gamma_{\nu\rho\delta}(-p_2,q_1,-q_2)\right] \times \frac{-i}{D_2} \left(g^{\delta \beta}-\frac{q_2^\delta q_2^\beta}{m_{2}^2} \right) \crn
&= 2e\,Q\, g_{hV_{ij}}\,g_{ZV_{ij}} \int \frac{d^dq}{(2 \pi)^d} \frac{1}{D_0D_1D_2}V_{1\mu\beta\lambda}V_{2\nu}^{\beta\lambda},
\end{align}
where $m_{1,2}\equiv m_{V_{i,j}}$, $D_0=q^2-m_1^2$, $D_{1,2}=q^2_{1,2}-m_2^2$,
\begin{align}
V_{1\mu\beta\lambda}&=g_{\alpha \beta}\left( g^{\alpha \alpha'} - \frac{q^\alpha q^{\alpha'}}{m_1^2} \right)\Gamma_{\mu\alpha'\lambda}(-p_1,q,-q_1),\crn
 V_{2\mu}^{\beta\lambda}&=\left( g^{\lambda \rho} - \frac{q_1^\lambda q_1^{\rho}}{m_{2}^2} \right)\times \left[\Gamma_{\nu\rho\delta}(-p_2,q_1,-q_2)\right] \left(g^{\delta \beta}-\frac{q_2^\delta q_2^\beta}{m_{2}^2} \right).\label{V12munu}
\end{align}
We note that factor 2 appearing in the first line of eq.~(\ref{M5Vij}) was added
in order to count two different diagrams with opposite internal lines in the loops.
It can be done because coupling constants $g_{hV_{ij}}$ and $g_{ZV_{ij}}$ are real numbers
in all models that we consider here.
Based on the structure of the PV functions, we know that $F_{21}p_{2\mu}p_{1\nu}$ gets
contributions from parts having the following factors: $q_{\mu}q_{\nu}$, $q_{\mu}p_{1\nu}$,
$p_{2\mu}q_{\nu}$, and $p_{2\mu}p_{1\nu}$. This means that we can do the following
replacements in the calculation:
\bea q_{1\mu}&\rightarrow& q_{\mu},\quad q_{2\mu}\rightarrow q_{\mu}-p_{2\mu},\quad q_{2\nu}\rightarrow q_{\nu}-p_{1\nu}=q_{1\nu},\crn
k_{1\mu}&\rightarrow& 0, \quad k_{2\mu}\rightarrow -p_{2\mu}, \quad k_{1\nu},\, k_{2\nu}\rightarrow -p_{1\nu},\quad g_{\mu\nu}\rightarrow0.
\label{ht}\eea

After some  intermediate steps shown in Appendix~\ref{detailsAmp}, and combining with  the relations $q^2=D_0+m_1^2$ and $D_{1,2}=q_{1,2}^2+m^2_2$, we have
\begin{align}
i\mathcal{M}_{(5)\mu\nu}&\rightarrow \left[e\,Q\, g_{hV_{ij}}\,g_{ZV_{ij}}\right]\times \int \frac{d^dq}{(2 \pi)^d} \times\frac{1}{m_1^2m_2^2}\crn
&\times \left\{q_{\mu}q_{\nu} \left[-\frac{1}{D_2} -\frac{1}{D_0} +\frac{2(m_1^2-m_2^2+m_Z^2)}{D_1D_2} +\frac{m_1^2+m_2^2+m_h^2)}{D_0D_2} \right.\right. \crn
&\left.\left. +\frac{8(d-2)m_1^2m_2^2 +2\left(m_1^2+m_2^2+m_h^2\right)(m_1^2+m_2^2-m_Z^2)}{D_0D_1D_2}\right]\right.\crn
&+\left. q_{\mu}p_{1\nu}\left[\frac{1}{D_2} +\frac{1}{D_0} -\frac{2(m_1^2-m_2^2+m_Z^2)}{D_1D_2} -\frac{5 m_1^2+3m_2^2 +m_h^2}{D_0D_2}  \right.\right. \crn
&\left.\left.+\frac{2(m_1^2+m_2^2-m_Z^2)}{D_0D_1} -\frac{8(d-2)m_1^2m_2^2 +2\left(m_1^2+m_2^2+m_h^2\right)(m_1^2+m_2^2-m_Z^2)}{D_0D_1D_2}\right]\right.\crn
&+\left.p_{2\mu}q_{\nu} \left[-\frac{4 m_1^2}{D_1D_2} +\frac{2m_1^2+4m_2^2}{D_0D_2} - \frac{4(m_1^2-m_2^2)(m_1^2+m_2^2-m_Z^2)}{D_0D_1D_2}\right]\right.\crn
&\left.+p_{2\mu}p_{1\nu} \left[\frac{4m_1^2}{D_1D_2} +\frac{2m_1^2}{D_0D_2} +\frac{4m_1^2(m_1^2 +3m_2^2 -m_Z^2)}{D_0D_1D_2}\right] \right\}.
\label{F21_5}
\end{align}
The calculation to derive the needed contribution from digram 6 in figure~\ref{loopHd} are the same way applied to diagram 1, see details in  Appendix~\ref{detailsAmp}.
Diagram 7  does not give any contributions. We can see that many divergent terms related to $q_{\mu}q_{\nu}$
in two amplitudes~(\ref{F21_5})  and~(\ref{iM6u1}) of diagram~6 will cancel out each other when
they are summed.  Hence, the pure gauge boson loops give the following total contribution:
 \begin{align}
 \mathcal{M}_{(5+6)\mu\nu}&\rightarrow e\,Q\, g_{hV_iV_j}\,g_{ZV_iV_j} \int \frac{d^dq}{(2 \pi)^d} \times\frac{1}{m_1^2m_2^2}\crn
 &\times \left\{q_{\mu}q_{\nu} \left[\frac{2(m_1^2-m_2^2+m_Z^2)}{D_1D_2} \right.\right. \crn
 &\left.\left. +\frac{8(d-2)m_1^2m_2^2 +2\left(m_1^2+m_2^2+m_h^2\right)(m_1^2+m_2^2-m_Z^2)}{D_0D_1D_2}\right]\right.\crn
 &+\left. q_{\mu}p_{1\nu}\left[\frac{1}{2D_2} +\frac{1}{2D_0} -\frac{2(m_1^2-m_2^2+m_Z^2)}{D_1D_2} -\frac{7(m_1^2+m_2^2) +m_h^2}{2D_0D_2}  \right.\right. \crn
 &\left.\left.+\frac{2(m_1^2+m_2^2-m_Z^2)}{D_0D_1} -\frac{8(d-2)m_1^2m_2^2 +2\left(m_1^2+m_2^2+m_h^2\right)(m_1^2+m_2^2-m_Z^2)}{D_0D_1D_2}\right]\right.\crn
 &+\left.p_{2\mu}q_{\nu} \left[ -\frac{1}{2D_2} -\frac{1}{2D_0}-\frac{4 m_1^2}{D_1D_2} +\frac{7(m_1^2+m_2^2) +m_h^2}{2D_0D_2}\right.\right. \crn
 &\left.\left. - \frac{4(m_1^2-m_2^2)(m_1^2+m_2^2-m_Z^2)}{D_0D_1D_2}\right]\right.\crn
 &\left.+p_{2\mu}p_{1\nu} \left[\frac{4m_1^2}{D_1D_2} +\frac{4m_1^2(m_1^2 +3m_2^2 -m_Z^2)}{D_0D_1D_2}\right] \right\}.
 \label{F21_56}
 \end{align}
Based on Appendix~\ref{Looptoolnote}, expression~(\ref{F21_56}) can be presented
explicitly in terms of the PV functions
$\mathcal{M}_{(5+6)\mu\nu}= \mathcal{M}_{(5+6)\mu\nu}(B_{0,\mu,\nu,\mu\nu}, C_{0,\mu,\nu,\mu\nu})\times 1/(16\pi^2)$. In addition, to keep only the parts with factor $p_{2\mu}p_{1\nu}$ we can
use the following replacements:
 \begin{align}
 &A^{(0)}_{\mu,\nu},\,A^{(1)}_{\mu},\,B^{(1)}_{\mu,\mu\nu} \rightarrow 0,\quad \left\{ A^{(2)}_{\mu},B^{(2)}_{\mu},\,B^{(12)}_{\mu}\right\} \rightarrow \left\{ A^{(2)}_0,-B^{(2)}_1,\frac{B^{(12)}_0}{2} \right\}p_{2\mu},\crn
 & A^{(1,2)}_{\nu},\,B^{(1,2)}_{\nu},\,B^{(12)}_{\nu}\rightarrow\left\{A^{(1,2)}_0,\,-B^{(1,2)}_1,\, B^{(12)}_0\right\}p_{1\nu},\quad  B^{(12)}_{\mu\nu}\rightarrow\frac{B^{(12)}_0}{2}p_{2\mu}p_{1\nu},\crn
&C_{\mu}\rightarrow -C_2\,p_{2\mu},\, C_{\nu}\rightarrow -(C_1+C_2)p_{1\nu}, \, C_{\mu\nu}\rightarrow (C_{12}+C_{22}) p_{2\mu} p_{1\nu}.
 \end{align}
Then, the total contribution from $V_i-V_{j}-V_j$ gauge boson loops is
\begin{align}
F_{21,V_{ijj}}&=\frac{2e\,Q\,g_{hV_{ij}}\,g_{ZV_{ij}}}{16\pi^2} \crn
&\times\left\{ \left[8+\frac{(m_1^2+m_2^2+m_h^2)(m_1^2+m_2^2-m_Z^2)}{m_1^2m_2^2}\right] \left(C_{12}+C_{22}+C_{2}\right)\right. \crn
&\left. +\frac{2(m_1^2-m_2^2)(m_1^2+m_2^2-m_Z^2)}{m_1^2m_2^2}(C_1+C_2) +\frac{2(m_1^2+ 3m_2^2-m_Z^2)C_0}{m_2^2} \right\},\label{F21Vijj}
\end{align}
where all PV functions having divergence completely disappeared, and therefore $d=4$.
We would like to emphasize now that formula~(\ref{F21Vijj}) is written
in terms of PV functions which are contained in LoopTools and hence it can be easily
evaluated numerically. Moreover, analytic expressions for the relevant PV functions
have been constructed~\cite{hzgamssm,C0f}, that is enough to implement our results
in existing numerical programs or to write a new stand-alone code.

We would like comment here about a more general case when couplings of gauge bosons
and photon do not obey the relation $g_{Z\gamma V_{ij}}=e\,Q\,g_{ZV_{ij}}$, which helps us
to reduce many divergent terms in $\mathcal{M}_{(5+6)\mu\nu}$. The key point here is that
the condition of on-shell photon always cancels out the most dangerous divergent terms
in the last line of~(\ref{V2_i}). As a by-product, the final form of $\mathcal{M}_{(5+6)\mu\nu}$
can contain more PV functions with divergent parts. Fortunately, all of them are
well-determined and widely used for numerical computation.

Before comparing our result with many well-known expressions computed in specific models,
we will introduce analytic  formulas for contributions from the remaining diagrams
listed in figure~\ref{loopHd} for completeness.

\subsection{Contributions from other diagrams in figure~\ref{loopHd}}

The contributions to $F_{21}$ from the first four diagrams in figure~\ref{loopHd} are
\begin{align}
F_{21,f_{ijj}}&=F^{(1)}_{21}=-\frac{e\,Q\,N_c}{16\pi^2}\left[ 4\left(K^+_{LL,RR} +K^+_{LR,RL} +\mathrm{c.c.}\right) \left(C_{12} +C_{22} +C_2\right) \right.
\crn &\left.\quad\quad\quad \quad+ 2\left(K^+_{LL,RR} -K^+_{LR,RL} +\mathrm{c.c.} \right) \left(C_1 +C_2\right) +2(K^+_{LL,RR} +\mathrm{c.c.})C_0\right],\crn
F_{5,f_{ijj}}&= -\frac{e\,Q\,N_c}{16\pi^2}\left[ 2\left(K^-_{LL,RR} -K^-_{LR,RL}- \mathrm{c.c }\right) \left(C_1 +C_2\right)-2(K^-_{LL,RR}- \mathrm{c.c.}) C_0\right], \label{F21fff}\\
F_{21,S_{ijj}}&=F^{(2)}_{21} =\frac{e\,Q\left(\lambda^*_{hS_{ij}}g_{ZS_{ij}} +\mathrm{c.c.}\right)}{16\pi^2} \left[ 4(C_{12}+C_{22} +C_2)\right],\label{F21sss}\\
%
F_{21,VSS}&=F^{(3)}_{21}=\frac{e\,Q\,(g^*_{hV_iS_j}g_{ZV_iS_j} +\mathrm{c.c.})}{16\pi^2}\crn
&\quad\quad\quad \quad\times \left[2 \left(1+\frac{-m_2^2+m_h^2}{m_1^2}\right)(C_{12}+C_{22} +C_2) +4(C_1+C_2 +C_0)\right],\label{F21VSS}\\
F_{21,SVV}&=F^{(4)}_{21}=\frac{e\,Q\, (g_{hV_jS_i}g^*_{ZV_jS_i} +\mathrm{c.c.})}{16\pi^2} \crn
& \quad\quad\quad \quad\times  \left[2 \left(1+\frac{-m_1^2+m_h^2}{m_2^2}\right)(C_{12}+C_{22} +C_2)-4(C_1+C_2)\right], \label{F21SVV}
\end{align}
where $m_{1,2}\equiv m_{X,Y}$ in the loop of  $F_{21,XYY}$,  $N_c$ is the colour factor coming from the $SU(3)_C$ symmetry, and the abbreviation
$\mathrm{c.c.}$ stands for the complex conjugated parts. The latter are the contributions
coming from diagrams having opposite directions of internal lines with respect to the ones
given in figure~\ref{loopHd}. Other relevant notations are
\begin{align}
\label{KLR}
K^{\pm}_{LL,RR}&=m_1\left(Y_{hf_{ij}L}\, g^*_{Zf_{ij}L}\pm  Y_{hf_{ij}R}\, g^*_{Zf_{ij}R}\right),\crn
K^{\pm}_{LR,RL}&=m_2\left(\pm Y_{hf_{ij}L}\, g^*_{Zf_{ij}R} + Y_{hf_{ij}R}\, g^*_{Zf_{ij}L}\right).
\end{align}
Details of calculating contributions from fermion loops $F_{21,f_{ijj}}$ are shown
in Appendix~\ref{detailsAmp}. Formulas for $F_{21,S_{ijj}}$ and $F_{21,VSS}$ are calculated
easily. The $F_{21,SVV}$ part was computed based on the result of $V_{2\mu}^{\beta\lambda}$
in eq.~(\ref{V1_12}). All steps we presented here were performed using the
FORM language~\cite{form1,form2}.

Formulas for $F_{21,f_{ijj}}$, $F_{5,f_{ijj}}$, and  $F_{21,S_{ijj}}$  are irrelevant for
the discussion of boson mediations. Similar general forms can be found in many previous
works, e.g., in~\cite{GMmodel, hdcay1, HzgaTHD}.  All of them are easy to check to be
consistent with our result so we will not present the comparison here.
We just focus on the most important formula $F_{21,V_{ijj}}$.

\section{\label{comparision}Comparison with previous results}
\subsection{The Standard Model}

The contribution of $W$ bosons corresponds to
$( g_{hV_{ij}},\,g_{ZV_{ij}},\,Q ) \rightarrow (g\,m_W,\, g\, c_W,\,1)$
with $m_1=m_2=m_W$, where $m_W$ is the $W$ boson mass, $g$ is the gauge coupling of
the $SU(2)_L$ group, $s_W\equiv \sin\theta_W$ with $\theta_W$ being the Weinberg angle.
Then formula~(\ref{F21Vijj}) is reduced to the simpler form:
\begin{align}
F^{\mathrm{SM}}_{21,W}&=\frac{e\,g^2 m_W \,c_W}{16\pi^2} \left\{ 2\left[ 8+ \left(2 +\frac{m_h^2}{m_W^2}\right)\left(2 -\frac{m_Z^2}{m_W^2}\right) \right] \left( C_{12} +C_{22} +C_2\right) + 4\left(4- \frac{m_Z^2}{m_W^2}\right) C_0\right\} \label{SMF21Wa}\\
&= \frac{\alpha_{\mathrm{em}}\,g\,c_W}{4\pi m_W\,s_W}  \left\{ \left[ 5+\frac{2}{t_2}- \left(1+ \frac{2}{t_2}\right)t^2_W   \right]I_{1}(t_2,t_1)- 4(3-t_W^2)I_2(t_2,t_1)\right\}, \label{SMF21W}
\end{align}
where we have used $\alpha_{\mathrm{em}}=e^2/(4\pi)$, $e=g\,s_W$, $m_h^2/m_W^2=4/t_2$,
$m_Z^2/m_W^2=4/t_1$, $m_Z^2/m_W^2=1/c^2_W=1+t_W^2$, $s_W=\sin{\theta_W}$, and $t_W=s_W/c_W$.
We also used the well-known functions $I_{1,2}(t_2,t_1)$ given in ref.~\cite{HHunter}
to identify $C_{12}+C_{22} +C_2= I_{1}(t_2,t_1)/(4m_W^2)$, and
$C_0=-I_2(t_2,t_1)/m_W^2$\footnote{The function $C_0$ in this special case is consistent
with the one from~\cite{HzgaTHD, GMmodel}, but different from the one in~\cite{hzgamssm}
by the opposite sign.}. They are proved in Appendix~\ref{specialf}. Formula~(\ref{SMF21W})
is consistent with well-known result for the SM case given in~\cite{HHunter,GMmodel}, which even has been confirmed using various approaches \cite{Boradjiev:2017khm}.

The right hand side of  eq.~(\ref{SMF21Wa}) can be proved to be completely consistent
with the $W$ contribution to the amplitude of the decay $h\rightarrow \gamma\gamma$
with $g_{ZWW}\rightarrow g_{\gamma WW}=e$, and in the limit $m_Z\rightarrow0$, equivalently
$t_1=4m_W^2/m_Z^2\rightarrow \infty$. The analytic form of this contribution is
known~\cite{HHunter, h2ga1}, namely
\begin{align}
F_{21,W}^{h\gamma\gamma,\mathrm{SM}}&=  \frac{\alpha_{\mathrm{em}}\,g}{4\pi m_W}\left[ 2+3t_2 + 3(2t_2-t_2^2) f(t_2)\right], \label{F21wh2gaSM}
\end{align}
where $t_2=4m_W^2/m_h^2$ and $f(x)$ is the well-known function given in
Appendix~\ref{specialf}. The partial decay width is
$\Gamma(h\rightarrow\gamma\gamma)=m_h^3/(64\pi)|F^{h\gamma\gamma,\mathrm{SM}}_{21}|^2$,
where  $F_{21}^{h\gamma\gamma,\mathrm{SM}}$ contains $F_{21,W}^{h\gamma\gamma,\mathrm{SM}}$.
The above determination of $F_{21,W}$ depends only on the diagrams with $W$ boson,
hence it should be the same in both cases of photon and $Z$ boson, except their masses
and couplings with the $W$ boson. For the case of photon we have
\begin{align}
C_0&=-\frac{1}{m_W^2} \lim_{t_1\rightarrow \infty}I_2(t_2,t_1)=-\frac{t_2f(t_2)}{2m_W^2}, \crn
C_{12}+C_{22}+C_2&=\frac{1}{4m_W^2} \lim_{t_1\rightarrow \infty}I_1(t_2,t_1)=\frac{1}{8m_W^2}\left[ -t_2 +t_2^2 f(t_2)\right], \label{I12photon}
\end{align}
where the expression for $C_0$ is the same as the one in~\cite{C0h2ga}.
By inserting two equalities~(\ref{I12photon}) into the right hand side of~(\ref{SMF21Wa})
with $m_Z=0$, we will obtain exactly eq.~(\ref{F21wh2gaSM}).

Regarding the fermionic contribution in the SM, we verify here the simple case of a single fermion without mixing and color factors, where
 $m_1=m_2=m_f$ and $Y_{hf_{ij}L}=Y_{hf_{ij}R}= e\, m_f/(2m_W\,s_W)$, leading to $$K^{+}_{LL,RR}=K^{+}_{LR,RL}=K^{+*}_{LL,RR}=K^{+*}_{LR,RL}= \frac{e}{2m_Ws_W}\times m^2_f (g_{ZfL} +g_{ZfR})$$ and $ K^{-}_{LL,RR}=K^{-}_{LR,RL}=K^{-*}_{LL,RR}=K^{-*}_{LR,RL}= m^2_f(g_{ZfL} -g_{ZfR})$. Two formulas (\ref{F21fff}) for fermionic contributions are
\begin{align}
F^{\mathrm{SM}}_{21,f}&=-\frac{e^2\,Q}{16\pi^2m_W s_W}\times m^2_f(g_{ZfL} +g_{ZfR})\left[ 8 \left(C_{12} +C_{22} +C_2\right) +2C_0\right]\crn
&=\frac{\alpha_{\mathrm{em}}\,g}{4\pi m_W}\times\left[\frac{-2Q\left(T^{3L}_f-2 Q s^2_W\right)}{s_W\,c_W} \right] \left[ I_{1}(t_2,t_1) -I_2(t_2,t_1)\right],
\crn F_{5,f_{ijj}}&=0, \label{F21fffSM}
\end{align}
where $g_{ZfL}+g_{ZfR}=\left(T^{3L}_f-2 Q s^2_W\right)\times g/c_W$, and $T^{3L}_f$ is the fermion weak isospin.
Formula~(\ref{F21fffSM}) coincides with the result given in~\cite{HHunter}.

At the one loop level, the effective coupling $h\gamma\gamma$ can be calculated using the 't Hooft - Feynman gauge \cite{Lavoura:2003xp}, which will be useful  to crosscheck with our result when the decay $h\rightarrow \gamma\gamma$ in a particular BSM is investigated.

\subsection{Recent results}

 The one-loop contribution from new gauge bosons in the GHU model was given in
ref.~\cite{hzgaGHU}, where the unitary gauge was mentioned without detailed explanations.
We see that the triple and quartic gauge boson couplings in this model also obey
the Feynman rules listed in table~\ref{tVcoupling}, hence our formula in
eq.~(\ref{F21Vijj}) is also valid. Because the final result in ref.~\cite{hzgaGHU} was
written in terms of only $B_0$ and $C_0$ functions, which are independent on the choice
of integration variable, it can be compared with our result. Translated into our
notation, the most important relevant part in ref.~\cite{hzgaGHU} is
\begin{align}
F^{\mathrm{GHU}}_{21,V}&= \left(m_1^4 +m_2^4 +10 m_1^2 m_2^2\right) E_+(m_1,m_2) + \left[ (m_1^2 +m_2^2) (m_h^2-m_Z^2) -m_h^2m_Z^2\right] E_{-}(m_1,m_2) \crn
&- \left[ 4 m_1^2m_2^2 (m_h^2-m_Z^2) +2m_Z^4(m_1^2+m_2^2)\right] \left(C_0 +C'_0\right),\label{FHGU21V}
\end{align}
where function $C'_0$ is determined by changing the roles of $m_1$ and $m_2$, and
\begin{align}
E_{\pm}(m_1,m_2)=1+\frac{m_Z^2}{m_h^2-m_Z^2}\left( B^{(2)}_0- B^{(1)}_0\right) \pm (m_2^2 C_0 +m_1^2 C_0'). \label{epm}
\end{align}
Formula~(\ref{FHGU21V}) should be equivalent to our result, namely to the sum
$F_{21,V_{ijj}}+ F_{21,V_{jii}}$. In the special case where $V_i\equiv V_j$, corresponding
to $m_1=m_2=m$, $C_0'=C_0= -I_2(t_2,t_1)/m^2$, and $C_{12} +C_{22}+C_2=I_1(t_2,t_1)/(4 m^2)$.
In fact we find the agreement between eq.~(3.18) of ref.~\cite{hzgaGHU} and our result, namely
\begin{align}
\delta F_{21}=\left. F^{\mathrm{GHU}}_{21,V}- \left[ \frac{16\pi^2}{2e\,Q\,g_{hV_{ij}}\,g_{ZV_{ij}}}( F_{21,V_{ijj}} +F_{21,V_{jii}})\right]\times \left[- m_1^2m_2^2(m_h^2-m_z^2)\right] \right|_{m_1=m_2}=0. \nonumber
\end{align}
But two general results are not the same, {\it i.e.} they differ by
$\delta F_{21}= -2 \left(m_1^2 C_0 + m_2^2 C_0'\right) m_Z^4$.

Except $F_{21,V_{ijj}}$ in eq.~(\ref{F21Vijj}), our formulas are  consistent with
the results given in ref.~\cite{GMmodel}, which were obtained by calculating
the decay amplitude of charged Higgs boson $h^{\pm}\rightarrow W^{\pm}\gamma$
in the 't~Hooft-Feynman gauge for the Georgi-Machacek model. In our notations,
$F_{21,S_{ijj}}$,  $F_{21,S_{i}VSS}$, and $F_{21,SVV}$ correspond to scalar,
vector-scalar-scalar, and scalar-vector-vector loop diagrams mentioned in
ref.~\cite{GMmodel}. By using the same notations from LoopTools, our results
and those of ref.~\cite{GMmodel} have the same form.

The consistency between our results and those in ref.~\cite{GMmodel} is explained
by the same Lorentz structures in couplings of the gauge bosons $Z$ and $W^{\pm}$.
An important difference is that the $W^{\pm}$ carry electric charges while the $Z$
does not. For a certain diagram with $W^+$ or $W^-$ in the final state, the directions
of internal lines are fixed, hence the complex conjugated terms are allowed in
the amplitude of the decay $h\rightarrow Z\gamma$, but not in that of
$H^{\pm} \rightarrow W^{\pm}\gamma$. Hence, except the pure gauge boson loop diagrams,
the  contributions to $h\rightarrow Z\gamma$ can be translated into those
to $H^\pm \rightarrow W^{\pm}\gamma$ by excluding all complex conjugated parts.
Of course, the mass $m_Z$ and couplings of the $Z$ boson must be replaced with those
of the $W^{\pm}$ bosons. This explanation can be checked directly based on our
calculations given above.

Regarding $F_{21,V_{ijj}}$, which presents the total vector loop contribution to
the decay amplitude $H^\pm\rightarrow W^{\pm}\gamma$, the explicit expression
derived from eq.~(\ref{F21Vijj}) reads
\begin{align}
F_{21,V_{ijj}}^{H^{\pm}W^{\pm}\gamma} &=\frac{e\,Q\,g_{hV_{ij}}\,g_{WV_{ij}}}{16\pi^2} \crn
& \times  \left\{ \left[8+\frac{(m_1^2+m_2^2+m_{H^{\pm}}^2)(m_1^2+m_2^2-m_W^2)}{m_1^2m_2^2}\right] \left(C_{12}+C_{22}+C_{2}\right)\right. \crn
&\left. +\frac{2(m_1^2-m_2^2)(m_1^2+m_2^2-m_W^2)}{m_1^2m_2^2}(C_1+C_2) +\frac{2(m_1^2+ 3m_2^2-m_W^2)C_0}{m_2^2} \right\},\label{F21Vijjw}
\end{align}
where $m_{H^\pm}$ is the charged Higgs boson mass, $g_{WV_{ij}}$ is the triple gauge
coupling of the $W$ boson, and $Q$ is always the electric charge of the gauge boson
$V_j$ coupling with the photon. We note that the factor $2$ in eq.~(\ref{F21Vijj})
is not counted anymore. Now, we only need to focus on the part generated by the loop
structures used to compare with the specific result given in~\cite{GMmodel}.
This case  corresponds to $m_1=m_Z,m_2=m_W=m_Zc_W$ and $m_{H^{\pm}}=m_5$ for
the decay $h^{+}_5\rightarrow W^{+} \gamma$. Formula~(\ref{F21Vijjw}) now has
the following form
\begin{align}
F_{21,V_{ijj}}^{H_5^{\pm} W^{\pm}\gamma} &\sim \left(9 +\frac{1}{c_W^2} +\frac{m_5^2}{m_W^2}\right) \left(C_{12}+C_{22}+C_{2}\right) +2\left(\frac{1}{c_W^2}-1\right)(C_1+C_2) +2\left(\frac{1}{c_W^2} +2 \right)C_0 \crn
&= 10(C_{12}+C_{22}+C_2) +6 C_0 + \frac{m_5^2}{m_W^2}(C_{12}+C_{22}+C_2)\crn
&+ \frac{s^2_W}{c^2_W}(C_{12} +C_{22} +2C_1 +3C_{2} +2 C_0),\label{F21Vijjwa}
\end{align}
which is different from the result given in ref.~\cite{GMmodel} by the coefficient $10$
instead of $12$ in front of the sum $(C_{12}+C_{22} +C_2)$.  We see that the two parts
in our result with coefficients $m^2_5/m_W^2$ and $s^2_W/c^2_W$ are consistent with
$S_{GGG}$ and $S_{XGG}$ in  ref.~\cite{GMmodel}, respectively. The difference in
the remaining part might arise due to a missed sign of the ghost contribution
$S_{\mathrm{ghost}}$.
%

An approach using Feynman gauge was introduced  in Ref.~\cite{Goodsell:2017pdq}, where the result must be implemented in some numerical packages. The results can be used to crosscheck with ours for consistence, but left for a further work.

\section{\label{application}Heavy charged boson effects on Higgs decays  $h\rightarrow Z\gamma$ in  BSM}

Because new heavy charged  gauge $V^{\pm}$ and Higgs bosons  $S^{\pm}$  appear in non-trivial gauge extensions of the SM, they may contribute to loop-induced SM-like Higgs decays $h\rightarrow \gamma\gamma$ and $h\rightarrow Z\gamma$. While the couplings  $h VV$ and $hSS$ consisting of  virtual identical  charged particles always contribute to both decay amplitudes, the couplings $hWV$ and $hWS$ of the SM-like Higgs boson only contribute to the  later.   These couplings may cause significant effects to Br$(h\rightarrow Z\gamma)$ in the light of  the very strict experimental constraints of   Br$(h\rightarrow\gamma\gamma)$ \cite{Khachatryan:2016vau}. When $m^2_{X}\gg m^2_{W}$ with $X=S,V$,  the loop structures of the form factors with at least one virtual $W$ boson have an  interesting property that $$F'_{WX}\equiv \left|\frac{F_{21,{WXX}} +F_{21,{XWW}}}{ eQg_{hXW}g_{ZXW}/(16\pi^2)}\right|\sim F'_{W}\equiv \left|\frac{F_{21,W} }{ eg_{hWW}g_{ZWW}/(16\pi^2)}\right| \sim  \mathcal{O} \left(\frac{1}{m_W^2}\right),$$ i.e., the same order with the $W$ loop contribution.

\begin{figure}[ht]
	\includegraphics[width=15cm]{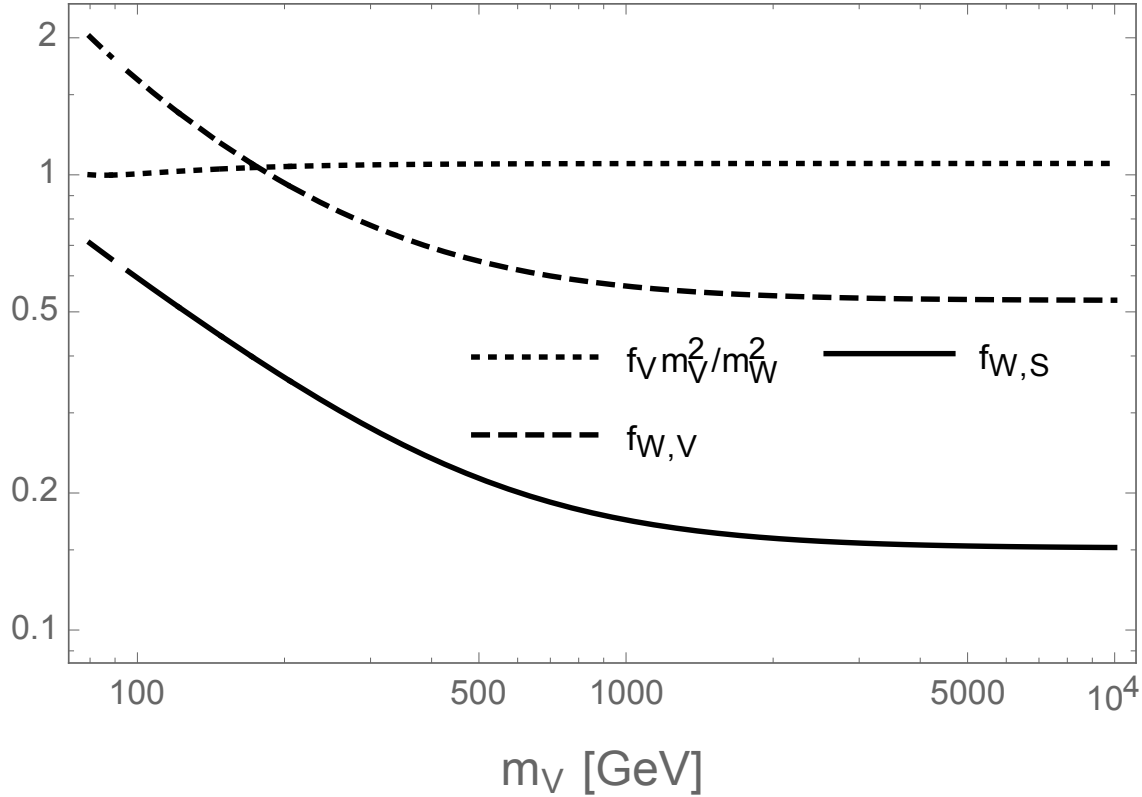}\\
	\caption{$f_V m_V^2/m_W^2$, $f_{W,S}$ and $f_{W,V}$ as functions of the $SU(2)_R$ scale $m_V$. }\label{fWX}
\end{figure}

 In contrast, the loop structure of a heavy gauge boson $F_{21,VVV}$ is
 $$F'_{V}\equiv \frac{F_{21,VVV}}{g_{hVV}g_{ZVV}/(16\pi^2)} \sim \mathcal{O}(m_V^{-2}),$$
  which is different from the SM contribution of the $W$ boson by  a factor $m^2_{W}/m_V^2$.  Numerical illustrations  are shown in figure~\ref{fWX} where  $f_{W,X}\equiv F'_{WX}/F'_{W}$, $f_{V}\equiv F'_{V}/F'_W$,  and $m_S=m_V$.
 Hence, the large coupling product $g_{hWX} g_{ZWX}$  may give significant effects on the total amplitude of the decay $h\rightarrow Z\gamma$.  But the contributions arising from this part were omitted in the literature, even with well known-models such as  the left-right models and the  Higgs Triplet Models (HTM).

 In the original LR models  reviewed in  \cite{gaugExtent221},  $g_{ZWW'}\sim (m_W/m_{W'})^2$,  lower bounds of few TeV for heavy gauge boson mass  $m_{W'}$ were concerned from recent experiments at LHC   \cite{Aaboud:2018vgh}.  As a result, its contributions  may be small. In contrast, recent  versions introducing different assignments of  fermions representations to explain latest experimental data of anomalies in $B$ meson decays allow lower values of $m_{W'}$ near 1 TeV \cite{lowmwp,Boucenna:2016qad}.

 Interesting studies on new charged gauge bosons $W'$ in left-right models \cite{Dobrescu:2015qna,Dobrescu:2015jvn,Dobrescu:2015yba}  indicated that the   couplings $W'Wh,\, W'WZ,W'H^\pm Z$ result in  important decays of  $W'^{\pm}$, which are being hunted at LHC.  These coupling also contribute to the decay $h\rightarrow Z\gamma$. The gauge bosons of the gauge groups $SU(2)_{L,R}$ and $U(1)_{B-L}$ are $W^a_{L,R\mu}$ ($a=1,2,3$) and $A_{B-L\mu}$~\cite{Dobrescu:2015jvn}, respectively. The Higgs sector consists of one bidoublet $\Sigma$ whose breaks the electroweak scale, and a $SU(2)_R$ multiplet whose breaks the $SU(2)_R\times U(1)_{B-L}$ scale.  Apart from the SM-like gauge bosons $Z_{\mu}$, $W^{\pm}_{\mu}$, and photon $A_{\mu}$, the left-right models predict new heavy gauge bosons including $W'^{\pm}$ and $Z'$ with  masses  $m_{W'}$ and $m_{Z'}$, respectively.  The bidoublet contributes mainly to the SM-like Higgs boson, Goldstone bosons of $Z$ and $W^{\pm}$, and a pair of singly charged Higgs $H^{\pm}$ that couple with the SM-like Higgs boson.

 Relevant vertex factors  are summarized  in Table~\ref{Wpcoup}. The details of the models and calculations are given in Appendix \ref{LRmodel}.
\begin{table}[ht]
	\begin{tabular}{ccc}
		\hline
		Vertex& SM &  LR~ \cite{Dobrescu:2015qna,Dobrescu:2015yba}\\
		\hline
		$g_{hWW}\, g_{ZWW}$&$g^2\,m_{W}\,c_W,$& $g_L^2\,m_{W}c_W\sin(\beta -\alpha)$ \\
		\hline
		$g_{hW'W}g_{ZWW'}$& $-$& $g_Lg_R\, m_W  \cos(\beta +\alpha) \frac{s_{\theta_+}}{c_W}$ \\
		\hline
		$g_{hW'W'}g_{ZW'W'}$&  $-$&   $-g^2_R\,m_{W}\,\sin(\beta -\alpha)  \frac{ s^2_W}{c_W}$\\
		\hline
		$g_{hW^+H^-}g_{ZW^-H^+}$&  $-$&   $-\frac{g^2_R}{2}\,m_{W}\,c_W\sin(\beta +\alpha)\cos(2\beta) s^2_{\theta_+}$\\
			\hline
		$g_{hW'^+H^-}g_{ZW'^-H^+}$&  $-$&   $-\frac{g^2_R}{2}\,m_{W}\,c_W\sin(\beta +\alpha)\cos(2\beta)$\\
		\hline	
	\end{tabular}
	\caption{Vertex factors involved charged gauge and Higgs bosons contributing to  one loop amplitude of the SM-like Higgs decay $h\rightarrow Z \gamma$ in the LR model with $g\equiv g_L$  and $s_{\theta_+}\simeq \tan{\theta_+}= \frac{g_R}{g_L}\times \sin(2\beta)\epsilon^2$, and $\epsilon=m_W/m_{W'}$.}\label{Wpcoup}
\end{table}
We have used the condition $\alpha=\beta-\pi/2$ to guarantee that the coupling $hWW$ is the same as that in the SM. We ignore all suppressed terms having factors with  orders larger than $\mathcal{O}(\epsilon^2)$, where $\epsilon=m_W/m_{W'}$ and $m_{W'}$ is  the new heavy gauge boson mass, which can be considered as the  breaking scale of the $SU(2)_R$ group. The couplings of the SM-like Higgs boson we discuss here are consistent with those in Ref. \cite{Dev:2016dja, Dobrescu:2015jvn, Dobrescu:2015yba, Jinaru:2013eya}. The triple gauge couplings are also consistent with Refs. \cite{gaugExtent221, Vijcouple}. Because they are not affected by the fermion assignments, they can be considered in the general case which does not depend on the recent experimental limit.

 With the above assumptions, the couplings of the SM-like Higgs boson are nearly the same as those in the  SM. The decay $h\rightarrow Z\gamma$ has contributions associated with charged gauge  bosons estimated as follows,
\begin{align} \label{WBoson_con}
\frac{F^{\mathrm{LR}}_{21,WWW}}{F^{\mathrm{SM}}_{21,W}}&\simeq 1,\quad
\frac{F^{\mathrm{LR}}_{21,W'W'W'}}{F^{\mathrm{SM}}_{21,W}}\sim  -\frac{g_R^2s^2_W}{g_L^2c^2_W}\epsilon^2,\crn
 \frac{F^{\mathrm{LR}}_{21,WW'W'} +F^{\mathrm{LR}}_{21,W'WW}}{F^{\mathrm{SM}}_{21,W}}&\sim  \frac{g_R^2\sin^2(2\beta)}{2g_L^2c^2_W}\epsilon^2, \quad
 \frac{F^{\mathrm{LR}}_{21,HW'W'} +F^{\mathrm{LR}}_{21,W'HH}}{F^{\mathrm{SM}}_{21,W}} \sim   \frac{g_R^2\cos^2(2\beta)}{2g_L^2}\epsilon^2,
\end{align}
 where $\epsilon\equiv m_W/m_{W'}$ and $\alpha\simeq \beta-\pi/2$. We can see that all quantities listed in (\ref{WBoson_con}) have the same order, although some of them are affected by the tiny mixing parameter $s_{\theta_+}= \mathcal{O}(\epsilon^2)$ between two charged gauge bosons. Hence all of them  must be taken into account. This argument is different from previous treatment where only $F^{\mathrm{LR}}_{21,W'W'W'}$ was mentioned \cite{hzgalr1, hzgalr2, Maiezza:2016bzp}. The recent lower bounds of the $SU(2)_R$ scale give $\epsilon^2\le  \mathcal{O}(10^{-3})$,   implying  that the  heavy charged Higg and gauge contributions discussed here are   suppressed. But the calculation  is very useful for further investigation in many other gauge extensions  allowing  lower new breaking scales, for example, the models belonging  to the class of  breaking pattern I mentioned in Ref. \cite{Vijcouple}, or recent models with breaking pattern II \cite{Boucenna:2016qad, lowmwp}.

The effects of heavy charged Higgs boson $m_{H}$ from  $F_{21,WSS}$ and  $F_{21,SWW}$  appear in  simple models like the HTM, for a review see \cite{Accomando:2006ga}. They even appear in the simple HTM models extended from the SM by adding  only one Higgs  triplet $\Delta$ ~\cite{Konetschny:1977bn,Cheng:1980qt,Schechter:1980gr}. It  contains one singly,  another doubly charged scalar components, and a neutral one with non-zero  expectation vacuum value (vev) denoted as $v_{\Delta}$. As a result, apart from the SM particles, the HTM predicts only new Higgs bosons. The   factors $g_{hSW}$ and $g_{ZWS}$  arise from couplings of singly charged Higgs  bosons $S^{\pm}$ with all gauge and neutral bosons.  The correlation of the two decays $h\rightarrow\gamma\gamma$ and $h\rightarrow Z\gamma$ were investigated previously, but the contributions $F_{21,WSS}$ and  $F_{21,SWW}$  mentioned here were ignored in  \cite{Dev:2013ff} because of the small product $g_{hSW}g_{ZWS}$.  It is proportional to the small  ratio $(v_{\Delta}/v)^2$ \cite{Arhrib:2011uy},  where  $v=246$ GeV. The requirement that the parameter $\rho=m_W^2/(m_Z^2c_W^2)$ is close to 1 at the tree level forces $v_{\Delta}$ to be  small with largest values  of  few GeV~\cite{Dev:2013ff, Aoki:2011pz,Blunier:2016peh}.  But the tree-level deviation $\Delta\rho=\rho-1$ predicted by this model is negative, in contrast with the   recent experimental results  \cite{Patrignani:2016xqp}. Hence, loop corrections should be included into this parameter, implying that  small $v_{\Delta}$ is no longer necessary  \cite{Gunion:1990dt, Accomando:2006ga}.  Theoretical prediction for $v_{\Delta}\sim \mathcal{O}(10)$ GeV is still allowed \cite{Kanemura:2012rs}.  The recent experimental  upper bound is $v_{\Delta}<25$ GeV \cite{Agrawal:2018pci}. As a result, contributions from $F_{21,SWW}$ and $F_{21,WSS}$ to the SM-like Higgs boson decay $h\rightarrow Z\gamma$ can reach value of $F_{21,W}\times \mathcal{O}(10^{-2})$, which is still far from the sensitivity of the recent experiments.  Hence, previous investigations ~\cite{Dev:2013ff, Arbabifar:2012bd, Aoki:2011pz} ignoring $F_{21,SWW}$ and $F_{21,WSS}$  in the one loop amplitude of the SM-like Higgs decay $h\rightarrow Z\gamma$ are still accepted.

  On the other hand,  heavy neutral bosons $H$  predicted by many BSM  may have large  $g_{HWS}\,g_{ZWS}$, for example the HTM   \cite{Arhrib:2011uy}. In this case,  contributions of $F_{21,SWW}$, $F_{21,WSS}$ can reach the significant values of  $F_{21,WWW}\times \mathcal{O}(v_{\Delta}/v)=F_{21,WWW}\times \mathcal{O}(10^{-1})$ in the decay  Br$(H\rightarrow Z\gamma)$ but they were ignored in previous works \cite{Arbabifar:2012bd,Chabab:2014ara, Blunier:2016peh}. The formulas  we introduced in this work  should be used for  improved  calculations of the mentioned decay rates.

\section{Conclusions}

The decay $h\rightarrow Z\gamma$ attracts now a great interest from both theoretical
and experimental sides. It should be observed and studied soon by the LHC experiments.
If a deviation from the SM prediction is found, it will be associated with new physics
implying additional contributions from exotic particles in many BSM models.
In this paper, we have introduced the general analytic formulas expressing one-loop
contributions from  scalars, fermions, and gauge bosons to the amplitude of
the decay $h\rightarrow Z\gamma$. In addition, we proved that our results can be used
to calculate the amplitude of the charged Higgs decays $H^{\pm} \rightarrow W^{\pm}\gamma$
which exist in many BSM models. Although some of these formulas were derived earlier
by other groups, the general forms were not concerned, in particular, the contributions
related to new gauge boson loops. Our formulas are applicable to many well-known gauge
extended versions of the SM, as we discussed in detail. We stress that all one-loop
contributions with gauge bosons involved are calculated explicitly using the unitary
gauge, so that the readers can cross-check our results. Our final results are written
in a convenient form. Namely, they are presented in terms of the standard
Passarino-Veltman functions which can be evaluated numerically with the help of
the LoopTools library. The analytic forms of these PV functions were also discussed,
so that our results can be identified with well known formulas in several special cases
as well as implemented into other numerical packages. Our results were checked to be
mainly consistent with several recent calculations in some specific BSM models, except the contributions from diagrams containing two different virtual gauge bosons.
We believe that our results will be useful for further studies of loop-induced decays
of neutral and charged Higgs bosons $H\rightarrow Z\gamma, W\gamma$, which have not
been yet treated in many well-known BSM models.

\section*{Acknowledgments}
L.T. Hue thanks Dr. LE Duc Ninh for enlightening discussions and comments about divergences and counterterms. He also thanks the BLTP, JINR for financial support and hospitality during his stay where this work is performed. The authors thank Prof.  Roberto Enrique Martinez and Dr.  Bhupal Dev, for communicating and recommending us Refs. \cite{Dev:2013ff,hzgalr1,hzgalr2}. This research is funded by the Vietnam National Foundation for Science and Technology Development (NAFOSTED) under the grant number 103.01-2017.29.

\appendix
\section{\label{Looptoolnote} PV functions in LoopTools}

\subsection{Definitions, notations and analytic formulas}

We use the notations for the Passarino-Veltman functions from the LoopTools
library~\cite{LoopTools}:
\begin{align}
A^{(i)}_{0,\mu}&=A_{0,\mu}(k_i^2; m_{i+1}^2)\equiv \frac{(2\pi\mu)^{4-d}}{i\pi^2} \int \frac{d^d q \,\left\{1,\, q_{\mu},\right\}}{D_i},\, i=0,1,2, \crn
B^{(i)}_{0,\,\mu,\,\mu\nu}&=B_{0,\mu}(k_i^2; m_1^2,m_{i+1}^2)\equiv \frac{(2\pi\mu)^{4-d}}{i\pi^2} \int \frac{d^d q \,\left\{1,\, q_{\mu},\,q_{\mu}q_{\nu}\right\}}{D_0D_i}, \, i=1,2,\crn
C_{0,\mu,\mu\nu}&=C_{0,\mu,\mu\nu}(p_1^2,p_2^2,(p_1+p_2)^2; m_1^2,m_2^2,m_3^2)\equiv \frac{(2\pi\mu)^{4-d}}{i\pi^2} \int \frac{d^d q \,\left\{1,\, q_{\mu},\, q_{\mu} q_{\nu}\right\}}{D_0D_1 D_2}, \label{PVlooptools}
\end{align}
where  $d=4-2\epsilon$ ($\epsilon\rightarrow0$) is the integral dimension,
$D_{i}=(q+k_{i})^2-m_{i+1}^2$,$k_0=0$, $k_1=-p_1$, $k_2=-(p_1+p_2)$,  $i=0,1,2$.
In our case, we always have $m_3=m_2$.

Denoting $\Delta_{\epsilon}=\frac{1}{\epsilon}+\ln(4\pi\mu^2)-\gamma_E$, it is well-known
that~\cite{peskin,aBfunc}
\begin{align}
A_0^{(0)}&=m_{1}^2(\Delta_{\epsilon}-\ln m_{1}^2+1),\quad A_0^{(1,2)}=m_{2}^2(\Delta_{\epsilon}-\ln m_{2}^2+1),\quad
%
%
A^{(i)}_{\mu} = - A^{(i)}_0k_{i\mu}, \quad
%
\label{Asfunc}
\end{align}
Based on the LoopTools notations~\cite{LoopTools}, functions $B^{(i)}_{0,\mu,\mu\nu}$
and $C_{0,\mu,\mu\nu}$ are written as
\begin{align}
B^{(i)}_{\mu}&= B^{(i)}_1 k_{i\mu}, \crn
 B^{(i)}_{\mu\nu}&= B^{(i)}_{00} g_{\mu\nu}+  B^{(i)}_{11} k_{i\mu}  k_{i\nu},\crn
 C_{\mu}&=C_1k_{1\mu} +C_{2}k_{2\mu},\crn
 C_{\mu\nu}&=C_{00} g_{\mu\nu} +C_{11}k_{1\mu}k_{1\nu} +C_{12}(k_{1\mu}k_{2\nu} +k_{2\mu}k_{1\nu}) +C_{22}k_{2\mu}k_{2\nu}.  \label{bcfunc}
\end{align}
There is another case where we have to change the integration variable
$q\rightarrow q'=q+k_1$ to get the standard form defined by~(\ref{PVlooptools}):
\begin{align}
B^{(12)}_{0,\mu,\mu\nu}&\equiv B_{0,\mu,\mu\nu}(k_1^2,k_2^2; m_2^2,m_2^2)= \frac{(2\pi\mu)^{4-d}}{i\pi^2} \int \frac{d^d q \,\left\{1,\, q_{\mu},\, q_{\mu} q_{\nu}\right\}}{D_1D_2}\crn
&=  \frac{(2\pi\mu)^{4-d}}{i\pi^2} \int \frac{d^d q \,\left\{1,\, q_{\mu}-k_{1\mu},\, (q_{\mu}-k_{1\mu})(q_{\nu}-k_{1\nu})\right\}}{(q^2-m_2^2)\left[(q+k_2-k_1)^2-m_2^2\right]}. \label{b12f}
\end{align}
Then we can use the scalar coefficients $B^{(12)}_0$, $B^{(12)}_{1}$, and $B^{(12)}_{11}$
with the standard definitions, where $k_2-k_1=-p_2$,
\begin{align}
B_{0,\mu,\mu\nu}((k_2-k_1)^2; m_2^2,m_2^2)&= \frac{(2\pi\mu)^{4-d}}{i\pi^2} \int \frac{d^d q \,\left\{1,\, q_{\mu},\, q_{\mu}q_{\nu}\right\}}{(q^2-m_2^2)\left[(q+k_2-k_1)^2-m_2^2\right]}, \crn
&= B^{(12)}_0,\,- B^{(12)}_1 p_{2\mu} ,\, B^{(12)}_{00} g_{\mu\nu}+  B^{(12)}_{11} p_{2\mu}p_{2\nu}. \label{bifuncp}
\end{align}
Inserting these into (\ref{b12f}) we get with $k_1=-p_1$ and $p_2^2=0$
\begin{align}
3B^{(12)}_{11}&=-2B^{(12)}_1=B^{(12)}_0=\Delta_{\epsilon}-\ln(m_2^2), \quad
B^{(12)}_{00}=\frac{m_2^2}{2}\left(1+B^{(12)}_0\right), \crn
B^{(12)}_{\mu}&=  \frac{B^{(12)}_0}{2}  p_{2\mu}+ B^{(12)}_0p_{1\mu},\crn
B^{(12)}_{\mu\nu}&= \frac{m_2^2}{2}\left(1+B^{(12)}_0\right) g_{\mu\nu}+ \frac{B^{(12)}_0}{3} p_{2\mu}p_{2\nu}+\frac{B^{(12)}_0}{2}\left( p_{2\mu}p_{1\nu} + p_{1\mu}p_{2\nu}\right)+ B^{(12)}_0 p_{1\mu}p_{1\nu}.
\end{align}
For two other cases we get
\begin{align}
B^{(i)}_{0}&\equiv B^{(12)}_0 +2-\sum_{\sigma=\pm}\left(1-\frac{1}{x_{i\sigma}}\right)\ln(1-x_{\sigma}),\crn
B^{(i)}_{1}&\equiv \frac{1}{2k_i^2}\left[A^{(0)}_0-A^{(i)}_0 -(m_1^2-m_2^2+k^2_i) B^{(i)}_0\right],
%
\end{align}
where $k_1^2=m_Z^2$, $k_2^2=m_h^2$, and $x_{i\sigma}$ are the roots of the equation
$m_2^2 x^2-(m_2^2-m_1^2+k_i^2)x +k_i^2 +i\epsilon=0$. The forms of $B^{(i)}_{0,1}$ used for
numerical investigation are well-known, see e.g.~\cite{aBfunc}.

The $C_0$ function with $m_3=m_2$ has a simple form~\cite{hzgamssm}:
\begin{equation}\label{C0f}
C_0=\frac{1}{k_1^2-k_2^2}\sum_{i=1}^2\sum_{\sigma=\pm}(-1)^i \mathrm{Li}_{2}\left[ \frac{2 k_i^2}{m_2^2-m_1^2+ k_i^2+\sigma \lambda^{1/2}(k_i^2,m_1^2,m_2^2)}\right],
\end{equation}
where $\lambda(x,y,z)=x^2+y^2+z^2-2xy-2yz-2xz$. This formula is also consistent with
LoopTools and~\cite{C0f}, where notations are changed as
$(m_1^2,m_2^2,m_F^2,m_B^2)\rightarrow (k_1^2,k_2^2,m_1^2,m_2^2)$.
The $C_{i,ij}$ functions are found based on the reduction technique~\cite{aBfunc}.
Their explicit forms used in this work are~\cite{C0f}
\begin{align}
C_1&=\frac{(m_h^2+m_Z^2)\left(B_0^{(1)}-B^{(12)}_0\right) -2m_h^2 \left(B_0^{(2)}- B_0^{(12)}\right)}{(m_h^2-m_Z^2)^2}+ \frac{f_2C_0}{m_Z^2-m_h^2},\crn
C_2&=\frac{(m_h^2+m_Z^2)\left(B_0^{(2)}-B^{(12)}_0\right) -2m_Z^2 \left(B_0^{(1)}- B_0^{(12)}\right)}{(m_h^2-m_Z^2)^2}- \frac{f_1C_0}{m_Z^2-m_h^2},\crn
C_{22}&=\frac{\left[ f_2\left(-3m_h^4+m_Z^4-4m_Z^2m_h^2\right)+ 4m_h^6-4m_Z^4m_h^2\right]B^{(2)}_0}{2 m_h^2\left(m_Z^2-m_h^2 \right)^3}
\crn&+\frac{3 f_1 m_Z^2 B^{(1)}_0}{(m_Z^2- m_h^2)^3} -\frac{\left[ f_1+f_2 +2(m_Z^2-m_h^2)\right] B^{(12)}_0}{2(m_Z^2-m_h^2)^2} +\frac{(f_1^2 +2 m_2^2 m_Z^2)C_0}{(m_Z^2-m_h^2)^2}\crn
&-\frac{(m_Z^2+m_h^2)\left(A^{(1)}_0 -A^{(0)}_0\right)}{2 m_h^2(m_Z^2-m_h^2)^2} +\frac{m_Z^2}{(m_Z^2-m_h^2)^2}
,\crn
C_{12}&=-\frac{\left[f_2(5 m_Z^2 +m_h^2) +m_Z^4-m_h^4 \right]B^{(1)}_0}{2(m_Z^2-m_h^2)^3} +\frac{\left[f_1(5 m_h^2 +m_Z^2) +m_h^4-m_Z^4 \right]B^{(2)}_0}{2(m_Z^2-m_h^2)^3}\crn
&+\frac{(2 m_2^2 -2 m_1^2 +m_Z^2 +m_h^2)B^{(12)}_0}{2 (m_Z^2 -m_h^2)^2} -\frac{\left[ f_1f_2 +m_2^2(m_Z^2 +m_h^2)\right]C_0}{(m_Z^2-m_h^2)^2}\crn
&+\frac{A^{(1)}_0 -A^{(0)}_0}{(m_Z^2-m_h^2)^2}  -\frac{m_Z^2 +m_h^2}{2(m_Z^2-m_h^2)^2},
\label{Cijgeneral}
\end{align}
where $f_i= m_2^2-m_1^2+k_i^2$. Some combinations which appear commonly in
our calculations are
\begin{align}
C_1+C_2&= -\frac{B^{(1)}_0-B^{(2)}_0}{m_Z^2-m_h^2}-C_0,\crn
C_{12}+C_{22}+C_2&= \frac{(-m_1^2 +m_2^2 +m_Z^2)(B^{(1)}_0-B^{(12)}_0)}{2(m_Z^2-m_h^2)^2}\crn
 &+\frac{\left[(m_1^2-m_2^2)(2m_h^2-m_Z^2)-m_Z^2m_h^2\right] (B^{(2)}_0 -B^{(12)}_0)}{2m_h^2(m_Z^2-m_h^2)^2} \crn
&+ \frac{m_2^2C_0}{m_Z^2-m_h^2} +\frac{m_1^2 -m_2^2+m_h^2-m_1^2\ln(m_1^2/m_2^2)}{2 m_h^2(m_Z^2-m_h^2)}.
\label{combi}
\end{align}

\subsection{\label{specialf}Analytic formulas in special case of $m_1=m_2=m$}

In the case of equal masses, we can use the following well-known
functions~\cite{hzgamssm, HzgaTHD, GMmodel}
\begin{align}
g(x)&=\left\{
\begin{array}{cc}
\sqrt{x-1}\arcsin\sqrt{\frac{1}{x}} & x \geq1, \\
\frac{\sqrt{1-x}}{2}\left(-i\pi+\ln\frac{1+\sqrt{1-x}}{1-\sqrt{1-x}}\right) & x<1 \\
\end{array}
\right.,
\label{gx}\\
f(x)&=\left\{
\begin{array}{cc}
\arcsin^2\sqrt{\frac{1}{x}} &x \ge1, \\
-\frac{1}{4}\left(-i\pi+\ln\frac{1+\sqrt{1-x}}{1-\sqrt{1-x}}\right)^2  & x<1 \\
\end{array}
\right.,
\label{fx}\\
I_1(x,y) &= \frac{xy}{2(x-y)}+\frac{x^2y^2}{2(x-y)^2} \left[ f(x)-f(y)\right] +\frac{x^2y}{(x-y)^2} \left[g(x)-g(y)\right], \label{I1xy}\\
I_2(x,y) &= -\frac{xy}{2(x-y)}\left[f(x)-f(y)\right]. \label{I2xy}
\end{align}
Defining  $t_1=t_z=4m^2/m_Z^2$ and $t_2=t_h=4m^2/m_h^2$, the PV functions involved
with this work can be written as
\begin{align}
B^{(i)}_0&= B^{(12)}_0+2- 2 g(t_i), \label{bi0m} \\
C_0&= -\frac{I_2(t_2,t_1)}{m^2},\label{C0m}\\
C_1+C_2&= \frac{B^{(1)}_0-B^{(2)}_0}{m_Z^2-m_h^2}-C_0,\crn
C_{12}+C_{22}+C_2&= \frac{  m_Z^2(B^{(1)}_0-B^{(2)}_0)}{2(m_Z^2-m_h^2)^2}+ \frac{m^2C_0}{m_Z^2-m_h^2} + \frac{1}{2 (m_Z^2 -m_h^2)}= \frac{I_1(t_2,t_1)}{4m^2}.
\end{align}
 The $B^{(i)}_0$ in eq. (\ref{bi0m}) is derived from the general well-know form, namely
 \begin{align}
 B^{(i)}_0&=B^{(12)}_0-\int_0^1 dx\ln\left[1 + 4t_i^{-1}x(x-1)\right]= B^{(12)}_0-\int_{-\frac{1}{2}}^{\frac{1}{2}} dx \ln\left[ 4t_i^{-1} x^2 + 1-t_i^{-1}\right].\nonumber
 \end{align}
More intermediate steps, including integration by parts, are as follows
 \begin{align}
 \int_{-\frac{1}{2}}^{\frac{1}{2}} dx \ln\left[ 4t_i^{-1} x^2 + 1-t_i^{-1}\right] &= -\int_{-\frac{1}{2}}^{\frac{1}{2}} \frac{8t_i^{-1} x^2\,dx}{4t_i^{-1} x^2 + 1-t_i^{-1}}=-2 + \int_{-\frac{1}{2}}^{\frac{1}{2}} \frac{2\,dx}{\frac{4x^2}{t_i-1} + 1}
 =-2+ 2 g(t_i).\nonumber
 \end{align}

 \section{\label{detailsAmp}Details of  amplitude calculation}
 For completeness, we present  some more detailed steps to obtain the formulas of   $M_{(5)\mu\nu}$ and  $M_{(5+6)\mu\nu}$ in Eqs.~\eqref{F21_5} and~\eqref{F21_56}. Also, the contribution from diagram 1 in figure~\ref{loopHd} will be discussed.

Using the replacements in Eq. \eqref{ht} to calculate the $F_{21}$ factor in $M_{(5)\mu\nu}$,   we get
 \begin{align}
 V_{1\mu\beta\lambda}
 &= g_{\alpha\beta}\left( g^{\alpha'}_\beta - \frac{q_\beta q^{\alpha'}}{m_1^2} \right) \left[ (q+q_1)_\mu g_{\alpha' \lambda} - (q+p_1)_\lambda g_{\alpha'\mu} - (q_1-p_1)_{\alpha'}g_{\mu \lambda} \right] \crn
 &=(q+q_1)_\mu g_{\beta \lambda} -(q+p_1)_\lambda g_{\mu \beta} -(q_1-p_1)_\beta g_{\mu \lambda} -\frac{(q+q_1)_\mu q_\beta q_\lambda}{m_1^2} \crn
 &+  \frac{(q+p_1)_\lambda q_\beta q_\mu}{m_1^2}+ \frac{(q_1-p_1)q q_\beta g_{\lambda \mu }}{m_1^2}\crn
 &\rightarrow  2 q_\mu g_{\beta \lambda} -(q+p_1)_\lambda g_{\mu \beta} -(q_1-p_1)_\beta g_{\mu \lambda} + \frac{q_\beta}{m_1^2}\left[-q_{\mu}q_{1\lambda}+ (q_1^2- m_Z^2)g_{\mu\lambda}\right]\crn
 &\equiv  V_{1,1\mu\beta\lambda}+ \frac{1}{m^2_{1}}\times V_{1,2\mu\beta\lambda}, \label{V1_12}
 \end{align}
 where we have used $p_1^2=m_Z^2$, $p_2^2=0$,
 $q.(q_1-p_1)=(q_1+p_1).(q_1-p_1)=q_1^2-p_1^2$, $q_1.(q_2-p_2)=q_2^2$ {\it etc.}
 The arrow means that replacements (\ref{ht}) have been applied.
 And we will apply them automatically from now on. Similarly, we can prove that
 \begin{equation}
 V_{2\nu}^{\beta\lambda}\rightarrow  V_{2,1\nu}^{\beta\lambda}+ \frac{1}{m^2_{2}}\times V_{2,2\nu}^{\beta\lambda}+ \frac{1}{m^4_{2}}\times V_{2,3\nu}^{\beta\lambda}, \label{V2_123}
 \end{equation}
 where
 \begin{align}
 V_{2,1\nu}^{\beta\lambda}&= -(q_1+p_2)^\beta  \delta_{ \nu}^\lambda -(q_2-p_2)^\lambda  \delta_{ \nu}^\beta +2 q_{1\nu} g^{ \beta \lambda },\crn
 V_{2,2\nu}^{\beta\lambda}&= q_1^{\lambda}\left(\delta^{\beta}_{\nu} q_2^2+ q_{1\nu}(q_1+p_2)^{\beta}-2q_{1\nu} q_{1}^{\beta}\right)+q_2^{\beta}\left(\delta^{\lambda}_{\nu} q_1^2- q_{2\nu}p_2^{\lambda}-q_{1\nu} q_{2}^{\lambda}\right),\crn
 &\rightarrow q_1^{\lambda}\delta^{\beta}_{\nu} q_2^2 +q_2^{\beta}\delta^{\lambda}_{\nu} q_1^2- 2q_{1\nu}q_1^{\lambda}q_{2}^{\beta},\crn
 V_{2,3\nu}^{\beta\lambda}&= -(q_1.p_2) q_1^{\lambda}q_2^{\beta} p_{2\nu}\rightarrow 0. \label{V2_i}
 \end{align}
 Now the part we need is written as follows
 \begin{equation}\label{pro_V1V2}
 V_{1\mu\beta\lambda}V_{2\nu}^{\beta\lambda}\rightarrow (V_{1,1}V_{2,1})_{\mu\nu}+\frac{(V_{1,1}V_{2,2})_{\mu\nu}}{m_2^2} + \frac{(V_{1,2}V_{2,1})_{\mu\nu}}{m_1^2} +\frac{(V_{1,2}V_{2,2})_{\mu\nu}}{m_1^2m_2^2},
 \end{equation}
 where
 \begin{align}
 (V_{1,1}V_{2,1})_{\mu\nu}&= 2(2d-3) q_{\mu}q_{\nu}+ (-4d+7)q_{\mu}p_{1\nu}-p_{2\mu}q_{\nu}+5p_{2\mu}p_{1\nu},\crn
 (V_{1,1}V_{2,2})_{\mu\nu}&= q_{\mu}q_{\nu}\left[ q^2 +q_1^2 +m_h^2 -2m_Z^2\right] +q_{\mu}p_{1\nu}\left[-q^2 -3q_1^2 +q_2^2 -m_h^2 +2m_Z^2\right]\crn
 & +p_{2\mu}q_{\nu}\left[ -2q^2 +q_1^2 +2m_Z^2 \right]+p_{2\mu}p_{1\nu}\left[2q^2 +q_1^2 -2 m_Z^2\right],\crn
 (V_{1,2}V_{2,1})_{\mu\nu}&= q_{\mu}\left[ -q_{1\nu}(q.q_2)+ q_{\nu} q_2^2\right] + (q_1^2-m_Z^2)\left(q_{\mu}q_{\nu} -2q_{\mu}p_{1\nu} +2p_{2\mu}q_{\nu}\right)\crn
 &= q_{\mu}q_{\nu}\left[-\frac{q^2}{2}+q_1^2 +\frac{q_2^2}{2} +\frac{m_h^2-2m_Z^2}{2}\right]\crn
 &+q_{\mu}p_{1\nu}\left[\frac{q^2}{2} -2q_1^2 +\frac{q^2_2}{2} +\frac{-m_h^2 +4m_Z^2}{2}\right] +p_{2\mu}q_{\nu}\left[ 2q_1^2 -2m_Z^2\right],\crn
 (V_{1,2}V_{2,2})_{\mu\nu}&= -q_{\mu}q_{\nu} p_1^2 q_2^2+ q_{\mu}q_{1\nu}(q.q_2)\left[2p_1^2-q_1^2\right]\crn
 &= q_{\mu}q_{\nu} \left[-m_Z^2 q_2^2 +\frac{1}{2}(q^2+q_2^2-m_h^2)\left(-q_1^2 +2m_Z^2\right)\right]\crn
 &- q_{\mu}p_{1\nu}\frac{1}{2}(q^2+q_2^2-m_h^2)\left(-q_1^2 +2m_Z^2\right).
 \label{proV1V2ij}
 \end{align}
 From this, it is easy to derive the Eq. \eqref{F21_5}.

 The amplitude corresponding to diagram~6 from figure~\ref{loopHd} is
 \begin{align}
 \label{iM6u}
 i\mathcal{M}_{(6)\mu \nu} &= \int \frac{d^dq}{(2 \pi)^d} (ig_{hV_{ij}}\,g_{\alpha \beta}) \frac{-i}{D_0} \left( g^{\alpha \alpha'} - \frac{q^\alpha q^{\alpha'}}{m_1^2} \right) \crn
 &\times (-i e\,Q~g_{ZV_{ij}}) \left[ 2 g_{\mu\nu}g_{\alpha' \beta'}-g_{\mu\nu}g_{\alpha' \beta'}-g_{\mu\nu}g_{\alpha' \beta'} \right]  \frac{-i}{D_2} \left(g^{\beta\beta'}-\frac{q_2^{\beta} q_2^{\beta'}}{m_2^2} \right) \crn
 &\rightarrow \left[e\,Q\,g_{hV_{ij}}~g_{ZV_{ij}}\right] \times \int \frac{d^dq}{(2 \pi)^d} \frac{1}{D_0D_2}\times\frac{1}{m_1^2m_2^2}\crn
 &\times \left[-2m_1^2q_{2\mu}q_{2\nu}-2m_2^2 q_{\mu}q_{\nu} +(q.q_2)(q_{2\mu}q_{\nu} +q_{\mu}q_{2\nu})\right].
 \end{align}
 Then it is easy to derive that
 \begin{align}
 i\mathcal{M}_{(6)\mu \nu}&\rightarrow  \left[e\,Q\,g_{hV_{ij}}~g_{ZV_{ij}}\right] \int \frac{d^dq}{(2 \pi)^d}\times \frac{1}{m_1^2m_2^2}\crn
 &\times\left\{ q_{\mu}q_{\nu}\left[\frac{1}{D_2} +\frac{1}{D_0} -\frac{m_1^2+m_2^2+m_h^2}{D_0D_2} \right]
 + q_{\mu}p_{1\nu}\left[-\frac{1}{2D_2}-\frac{1}{2D_0} +\frac{3m_1^2-m_2^2+m_h^2}{2D_0D_2} \right]\right.\crn
 &\left.+ p_{2\mu}q_{\nu}\left[-\frac{1}{2D_2}-\frac{1}{2D_0} +\frac{3m_1^2-m_2^2+m_h^2}{2D_0D_2} \right] +p_{2\mu}p_{1\nu}\left[\frac{-2m_1^2}{D_0D_2}\right]\right\}. \label{iM6u1}
 \end{align}
 Contribution from the diagram 1 of figure~\ref{loopHd} is
 \begin{align}
 i\mathcal{M}_{(1)\mu\nu}&=(-1)\times \int \frac{d^dq}{(2 \pi)^d}\times  \mathrm{Tr}\left[-i\left(Y_{hf_{ij}L}\,P_L +Y_{hf_{ij}R}\,P_R\right) \frac{i(q\!\!\!/_2+m_2)}{D_2}\right. \crn
 &\left. \times (ie\,Q\,\gamma_{\nu})\frac{i(q\!\!\!/_1+m_2)}{D_1}\left[ i\left(g^*_{Zf_{ij}L}\gamma_{\mu}\,P_L +g^*_{Zf_{ij}R}\gamma_{\mu}\,P_R\right)\right] \frac{i(q\!\!\!/+m_1)}{D_0}\right]\crn
 &= -e\,Q \int \frac{d^dq}{(2 \pi)^d}\times \frac{1}{D_0D_1D_2}
 \crn&\times\frac{1}{2} \mathrm{Tr}\left[ \left(q\!\!\!/_2\gamma_{\nu}q\!\!\!/_1\gamma_{\mu} + m^2_2\gamma_{\mu}\gamma_{\nu}\right) \left( K^{+}_{LL,RR} -K^{-}_{LL,RR}\gamma_5\right)\right.
 \crn&\left.+ \left(q\!\!\!/_2\gamma_{\nu}\gamma_{\mu}q\!\!\!/ +\gamma_{\nu}q\!\!\!/_1\gamma_{\mu}q\!\!\!/ \right) \left( K^{+}_{LR,RL} +K^{-}_{LR,RL}\gamma_5\right)\right].\nonumber
 \end{align}
 While the contribution of the corresponding diagram with opposite internal directions is
 \begin{align}
 i\mathcal{M}'_{(1)\mu\nu}&= -e\,Q \int \frac{d^dq}{(2 \pi)^d}\times \frac{1}{D_0D_1D_2}
 \crn&\times\frac{1}{2} \mathrm{Tr}\left[ \left(q\!\!\!/_2\gamma_{\nu}q\!\!\!/_1\gamma_{\mu} + m^2_2\gamma_{\mu}\gamma_{\nu}\right) \left(K^{+*}_{LL,RR}  +K^{-*}_{LL,RR}\gamma_5\right)\right.
 \crn&\left.+ \left(q\!\!\!/_2\gamma_{\nu}\gamma_{\mu}q\!\!\!/ +\gamma_{\nu}q\!\!\!/_1\gamma_{\mu}q\!\!\!/ \right) \left( K^{+*}_{LR,RL} -K^{-*}_{LR,RL}\gamma_5\right)\right].\nonumber
 \end{align}
 The sum of the two above diagrams gives the final result of $F_{21,f_{ijj}}$ and
 $f_{5,f_{ijj}}$ where the complex conjugation corresponds to the contribution
 from $\mathcal{M}'_{(1)\mu\nu}$. Using the properties of the Dirac matrices,
 it is easy to find out the two expresions given in eq.~(\ref{F21fff}).

 \section{\label{LRmodel}Gauge bosons and  couplings in the  left-right model $SU(2)_L\times SU(2)_R\times U(1)_{B-L}$}
 The model used here was introduced in Ref. ~\cite{Dobrescu:2015jvn,Dobrescu:2015yba}, where many results we show here were introduced.  The relations between the original gauge boson states   and the physical ones $\{W'^{\pm}_{\mu},\,W^{\pm}_{\mu}, A_{\mu},\, Z_{\mu},\,Z'_{\mu}  \}$  are
 \begin{align} \label{g_lrmodel}
 \begin{pmatrix}
 W^{\pm}_{R\mu}\\
 W^{\pm}_{L\mu}
 \end{pmatrix}&= \begin{pmatrix}
 c_{\theta_+}& s_{\theta_+}  \\
 -s_{\theta_+}& c_{\theta_+}
 \end{pmatrix} \begin{pmatrix}
 W'^{\pm}_{\mu}\\
 W^{\pm}_{\mu}
 \end{pmatrix}, \crn
 \begin{pmatrix}
 W^3_{L\mu}\\
 W^3_{R\mu}\\
 A_{B-L\mu}\\
 \end{pmatrix} &\simeq \begin{pmatrix}
 s_W, & c_W,  & -c^3_R\frac{g_R}{g_L}\epsilon^2 \\
 s_Rc_W, & -s_R s_W,  &c_R \\
 c_Rc_W, &-c_Rs_W,    & -s_R
 \end{pmatrix}\begin{pmatrix}
 A_{\mu}\\
 Z_{\mu}\\
 Z'_{\mu}\\
 \end{pmatrix} ,
 \end{align}
 where  $W^{\pm}_{L,R\mu}\equiv \frac{W^1_{L,R\mu}\mp i W^2_{L,R\mu}}{\sqrt{2}}$,
 \begin{align*}
 s_{\theta_+}=\frac{g_R}{g}\epsilon^2 \sin2\beta,\quad s_R&\equiv \frac{g_Y}{g_R}= \frac{g_L t_W}{g_R},\quad \epsilon\equiv \frac{m_W}{m_{W'}},  \quad \quad m_{Z'}=\frac{m_{W'}}{c_R}.
 \end{align*}
We will keep the approximation up to the order $\mathcal{O}(\epsilon^2)$, which gives $s^2_{\theta_+}=0$ and $c_{\theta_+}=1$.

 Only the bidoublet Higgs $\Sigma\sim(2,2,0)$ contributes to the SM-like Higgs boson, namely  \begin{align} \label{biHiggs}
 \Sigma=\begin{pmatrix}
 \Sigma^0_1&  \Sigma^+_2\\
 \Sigma^-_1& \Sigma^0_2
 \end{pmatrix} = \begin{pmatrix} v_H c_{\beta}- \frac{s_{\alpha}}{\sqrt{2}} h,
 & H^+c_{\beta}  \\
 H^-s_{\beta}, & v_H s_{\beta} +\frac{c_{\alpha}}{\sqrt{2}} h
 \end{pmatrix},
 \end{align}
 where only the SM-like Higgs $h$ and charged Higgs bosons are kept. The SM-like gauge boson $W^{\pm}$ has mass $m_W\simeq g_L v_H/\sqrt{2}$.

 The respective covariant derivative is \cite{gaugExtent221},
 \begin{align}
 \label{Dsigam}
 D_{\mu}\Sigma &= \partial_{\mu}\Sigma -ig_L \frac{\sigma_a}{2} W^a_{L\mu}\Sigma +ig_R\Sigma\frac{\sigma_a}{2} W^a_{R\mu} ,\crn
 &\equiv \partial_{\mu}\Sigma -\frac{ig_L}{2} P_{\Sigma\mu},
 \end{align}
 where $g_{L,R}$ and $W^a_{L,R\mu}$ $(a=1,2,3)$ are  the gauge couplings and bosons of the groups $SU(2)_{L,R}$, $\sigma_a$ are Pauli matrices.

 The kinetic term of the $\Sigma$ is
 \begin{align}\label{lkSigam}
 \mathcal{L}_{\Sigma}^k &= \mathrm{Tr}\left[\left( D_{\mu}\Sigma\right)^{\dagger } \left( D^{\mu}\Sigma\right)\right]\crn
 &=\mathrm{Tr}\left[ \partial_{\mu}\Sigma^{\dagger } \left( \partial^{\mu}\Sigma\right) -\frac{ig_L}{2} \left[ \partial_{\mu}\Sigma^{\dagger } \left(P^{\mu}\Sigma\right) - \left(P_{\mu}\Sigma\right)^{\dagger} (\partial^{\mu}\Sigma)\right] + \frac{g^2_L}{4}\left(P_{\mu}\Sigma\right)^{\dagger}\left(P^{\mu}\Sigma\right)\right],
 \end{align}
 which contains couplings of Higgs and gauge bosons. The part of the Lagrangian \eqref{lkSigam}  giving  couplings  $hV^+V'^-$ is
 \begin{align}
 \label{h0vpvm}
 \mathcal{L}(hV^{\pm}V'^{\mp})&= \frac{g^2_L}{2} \left[ \left( \Sigma^{0*}_1\Sigma^{0}_1 + \Sigma^{0*}_2\Sigma^{0}_2\right) \left( W^{+\mu}_L W^{-}_{L\mu} + \frac{g_R^2}{g_L^2}W^{+\mu}_R W^{-}_{R\mu}\right) \right. \crn
 &
 \left.- 2\frac{g_R}{g_L}\left( \Sigma^{0*}_1\Sigma^{0}_2 W^{+\mu}_L W^{-}_{R\mu}+\Sigma^{0*}_2 \Sigma^{0}_1 W^{+\mu}_R W^{-}_{L\mu}\right)      \right]\crn
 \rightarrow &  g_L m_W  \sin(\beta-\alpha) h \left( W^{\mu} W^{-}_{\mu} + \frac{g_R^2}{g_L^2}W'^{+\mu} W'^{-}_{\mu}\right) \crn
 & -g_R m_W  \cos(\beta+\alpha) h \left( W^{+\mu} W'^{-}_{\mu} + W^{-\mu} W'^{+}_{\mu}\right),
 \end{align}
 where we keep only dominant contributions to the coefficients of the $hV^+V'^-$, i.e. we  use the approximation $W\simeq W_L$ and $W'\simeq W_R$.

 The couplings  $ZH^\pm V^{\mp}$ are
 \begin{align} \label{zhv}
 \mathcal{L}(ZH^{\pm}V^{\mp}) &=-g_R m_W  \cos(2\beta) \times W^3_{L\mu} (W^{+\mu}_{R}H^- +W^{-\mu}_{R}H^+) \crn
 &\simeq -g_R c_W m_W  \cos(2\beta) \times Z_{\mu} \left( s_{\theta_+}W^{+\mu}H^- + W'^{+\mu}H^-+ \mathrm{H.c.}\right),
 \end{align}
 where we used $c_{\theta_+}=1$ and $W^{3\mu}_L \rightarrow c_W Z^{\mu}$. This result is consistent with~\cite{Dobrescu:2015jvn}

 The couplings $hH^\pm V^{\mp}$ are   %
 \begin{align}\label{h0HV}
 \mathcal{L}(hH^{\pm}V^{\mp})&= -\frac{ig_L}{2}\mathrm{Tr} \left[ \partial_{\mu}\Sigma^{\dagger } \left(P^{\mu}\Sigma\right) - \left(P_{\mu}\Sigma\right)^{\dagger} (\partial^{\mu}\Sigma)\right]\crn
  &\rightarrow \frac{g_L}{2}\cos(\beta -\alpha)\left[ (p_0-p_-)_{\mu} W^{+\mu}_LH^-h-  (p_0-p_+)_{\mu} W^{-\mu}_LH^+h \right]\crn
 &+\frac{g_R}{2}\sin(\beta +\alpha)\left[  (p_0-p_-)_{\mu} W^{+\mu}_RH^-h-  (p_0-p_+)_{\mu} W^{ -\mu}_RH^+h\right],
 \end{align}
 where we have used $\partial_{\mu}\rightarrow -ip_{\mu}$; $p_{0,\pm}$  are momenta of the Higgs boson $h$ and $H^{\pm}$. The first line of the final result in \eqref{h0HV} contains the factor $\cos(\beta-\alpha)\simeq \cos(\pi/2)=0$, because the matching condition with the SM coupling $hW^+W^-$ lead to $\beta=\alpha +\pi/2$. Using $W^{\pm}_{R\mu}\simeq W'^{\pm}_{\mu} +s_{\theta}W^{\pm}_{\mu}$, the second line is written in the physical gauge boson states as follows,
 \begin{align}\label{h0HV1}
 \mathcal{L}(hH^{\pm}V^{\mp})&= \frac{g_R}{2}\sin(\beta +\alpha)\left[  (p_0-p_-)_{\mu} \left( W'^{+\mu} +s_{\theta}W^{+\mu}\right)H^-h \right. \crn
 &\left. -
  (p_0-p_+)_{\mu}\left( W'^{-\mu} +s_{\theta}W^{-\mu}\right)H^+h\right].
 \end{align}
 The triplet couplings of three gauge bosons $ZVV'$ are contained in the kinetic term of the non-abelian gauge bosons, namely  \cite{gaugExtent221}
 \begin{align} \label{lkg}
 \mathcal{L}^k_g&=-\frac{1}{4}F^a_{L\mu\nu}F^{a\mu\nu}_L -\frac{1}{4}F^a_{R\mu\nu}F^{a\mu\nu}_R,\crn
 F^a_{L,R\mu\nu}&= \partial_{\mu} W^a_{L,R\nu}-\partial_{\nu} W^a_{L,R\mu} +  g_{L,R} \epsilon^{abc}W^b_{L,R\mu}W^c_{L,R \nu}.
 \end{align}
 The triplet gauge couplings are derived as follows,
 \begin{align} \label{l3g}
 \mathcal{L}_{3g}&=- g_{L} \epsilon^{abc} (\partial_{\mu} W^a_{L\nu})W^{b\mu}_LW^{c\nu}_L -g_{R} \epsilon^{abc} (\partial_{\mu} W^a_{R\nu})W^{b\mu}_RW^{c \nu}_R \crn
 &= -ig_L c_W \left[Z^{\nu} \left( -\partial_{\mu}W^+_{L\nu} W^{-\mu}_L + \partial_{\mu}W^-_{L\nu} W^{+\mu}_L\right)  +Z^{\mu} \left( \partial_{\mu}W^+_{L\nu} W^{-\nu}_L - \partial_{\mu}W^-_{L\nu} W^{+\nu}_L\right) \right. \crn
 &\left. +\partial_{\mu}Z_{\nu} \left( -W^{+\mu}_L W^{-\nu}_L + W^{-\mu}_L W^{+\nu}_L\right)  \right] -ig_R(-s_Rs_W)\times  (L\rightarrow ~R),
 \end{align}
 where we pay attention to only $Z$ couplings by replaced $W^3_L\rightarrow c_W Z$ and  $W^3_R\rightarrow -s_Rs_W Z$ in the last row of \eqref{l3g}.

 Now based on the Feynman rules,  the vertex factor of the coupling $Z^{\alpha}W^{+\mu}W^{-\nu}$ defined as $-ig_{ZW^+W^-}\Gamma_{\alpha\mu\nu}(p_0,p_+,p_-)$ can be derived by taking the limit $W^{\pm}_{L}\rightarrow W^{\pm}$. As a result, we obtain $g_{ZW^+W^-}\simeq g_Lc_W$.  Similarly, the coupling $Z^{\alpha}W'^{+\mu}W'^{-\nu}$ with the vertex factor $-ig_{ZW'^+W'^-}\Gamma_{\alpha\mu\nu}(p_0,p_+,p_-)$ gives  $g_{ZW'^+W'^-}\simeq -g_Rs_Rs_W=-g_Ys_W=-g_L s^2_W/c_W$.

 Using $W^+_{L\mu }W^-_{L\nu }\rightarrow -s_{\theta_+}c_{\theta_+} W'^+_{\mu }W^-_{\nu } +\mathrm{H.c.}$ and $W^+_{R\mu }W^-_{R\nu }\rightarrow s_{\theta_+}c_{\theta_+} W'^+_{\mu }W^-_{\nu } +\mathrm{H.c.}$,  the couplings  $Z^{\alpha}W'^{+\mu}W^{-\nu}$ and $Z^{\alpha}W^{+\mu}W'^{-\nu}$  give   $g_{ZW^+W'^-}=g_{ZW'^+W^-}=-s_{\theta_+}c_{\theta_+}\left( g_Lc_W + g_Rs_Rs_W\right)\simeq -g_Ls_{\theta_+}/c_W $, respectively.

\end{document}